# Asymmetric Independence Model for Detecting Interactions between Variables

Guoqiang Yu*

*Dept. of Electrical and Computer Engineering, The Virginia Polytechnic Institute and State University, Arlington, VA, USA.*

David J. Miller

*Dept. of Electrical Engineering, The Pennsylvania State University, State College, PA, USA.*

Carl D. Langefeld

*Dept. of Biostatistical Sciences, Wake Forest University, Winston-Salem, NC, USA.*

David M. Herrington

*Dept. of Internal Medicine, Wake Forest University, Winston-Salem, NC, USA.*

Yue Wang

*Dept. of Electrical and Computer Engineering, The Virginia Polytechnic Institute and State University, Arlington, VA, USA.*

**Summary.** Detecting complex interactions among risk factors in case-control studies is a fundamental task in clinical and population research. However, while hypothesis testing using logistic regression (LR) is a convenient solution, the LR framework is poorly powered and ill-suited under several commonly occurring circumstances, including missing or unmeasured risk factors, imperfectly correlated "surrogates", and multiple disease sub-types. The weakness of LR in these settings stems from the way the null hypothesis is defined. Here we propose the Asymmetric Independence Model (AIM) as a biologically-inspired alternative to LR, based on the key observation that the mechanisms associated with acquiring a "disease" versus maintaining "health" are asymmetric. We prove mathematically that, unlike LR, AIM is a robust model under the abovementioned confounding scenarios. Further, we provide a mathematical definition of a "synergistic" interaction, and prove theoretically that AIM has better power than LR for such interactions. We then experimentally show the superior performance of AIM compared to LR on both simulations and four real datasets. While the principal application here involves genetic or environmental variables in the life sciences, our methodology is readily applied to other types of measurements and inferences, *e.g.* in the social sciences.

---

*to whom correspondence should be addressed

# 1 Introduction

The problem of detecting statistical interactions between multiple factors that influence disease risk, treatment efficacy, or other complex human traits is pervasive in genetics, epidemiology, pharmacology, and social sciences, e.g. (Martinelli(1999)), (Castellsague(1999)), (Mallal(2008)), (Rosing(1999)), (Orkunoglu-Suer(2008)). In genetic studies, including candidate gene studies and genome-wide association studies (GWAS), once main effects are detected, finding gene-gene and gene-environment interactions may give crucial clues to the underlying biological mechanisms/pathways in play ( Matsuo(2001)), (Braun(2011)). Detection of interactions may also help to identify subpopulations that would benefit from, or be harmed by a particular drug treatment/intervention option, e.g. the effect of the anti-HIV drug abacavir in individuals with the HLA-B*5701 allele (Mallal(2008)). More generally, in the life, social, and physical sciences, detection of interaction effects can help generate new hypotheses or illuminate novel etiologic mechanisms. In all of these settings a statistical interaction is in general present when, as measured over a finite population, the joint effect of multiple risk factors or predictors differs, in a statistically significant fashion, from the expected baseline joint effect that would occur if these factors act independently. Inference on the presence of such interactions amongst a set of target factors should not be confused with the related, albeit quite distinct task of detecting novel risk factors by incorporating both interaction effects and (possibly weak) main effects (Marchini(2005)),(Cordell(2009)).

While detecting interactions is ubiquitous, the technical challenge associated with this inference task is very different from that associated with other statistical inference tasks. To elaborate, correctness of hypothesis testing rests heavily on the null hypothesis, while efficiency or power is often determined by the alternative hypothesis. For many inference tasks, the proper choice of the null model is plainly obvious and methodological development therefore focuses on tuning the alternative model to make it as sensitive as possible to enhance detection power. However, when testing for an interaction, careful specification of the alternative hypothesis is much less important than the null for the following reason. Although there are potentially different forms of the alternative hypothesis, they are all essentially the same in that they all involve fully saturated models. For instance, to detect an interaction effect that involves two binary variables, at least four parameters are necessary for the alternative hypothesis (modeling a main effect with two variables requires three parameters, one for each variable plus one baseline parameter, with the interaction effect requiring at least one more parameter). Recalling that two binary variables can specify at most four combinations, a model with four parameters is saturated (Hosmer(2013)), with all saturated models producing the same likelihood on given data. In contrast, in the setting of a possible interaction, different null models are not fully saturated and will not in general produce the same data likelihood. Thus, we argue that, for detecting interactions between factors, the null model must be carefully designed.

The null is usually called an independence model, describing how multiple factors jointly determine an outcome when there is no interaction between them. In this work, the focus will be on biological domains, where the outcome is usually a phenotype such as disease or healthy status. There are multiple non-equivalent ways to define this so-called baseline (null) joint effect. For example, in epidemiology, if we consider each disease factor to act independently, one model for baseline disease risk given the presence of multiple factors is the sum of the individual risks coming from each factor, i.e., an additive model. Alternatively, one can posit that the risk from multiple independent factors should be the product of their individual relative risks, i.e., a multiplicative baseline model (Gonzalez(2005)). Several other baseline models that have been previously proposed are reviewed in the Discussion section; however, they are all variants of these two principal models. Since these models are all mathematically valid, with each describing one class of possible baseline effects, it may appear that there is no objective basis for preferring one baseline effects model over another. However, we argue such basis can be found in the relationship both to existing knowledge and to practical considerations: 1) consistency of the model with existing theories/knowledge of the domain (where, here, the focus will be on biological domains involving a "disease status" phenotype); 2) any useful theoretical properties that accrue to the model and bear on its suitability in practice. As will be explained shortly, it turns out that the existing models have severe limitations undermining these two principles. These limitations are not only conceptual and theoretical. Significant implications such as inflated type I error and reduced power are observed and will be illustrated in the sequel. All of this inspired us to develop an alternative baseline effects model. To help motivate this new model, it is instructive to first consider for comparison the typical LR framework that is both familiar and widely applied.

**Logistic Regression for Interaction Detection**

Suppose there are $N$ binary factors, i.e. $\underline{X} = (X_1, X_2, \ldots, X_N)$, $X_i \in \{0, 1\}$, a binary health status variable $C \in \{0, 1\}$ ('case' = 1, 'control' = 0) and let $P(C = c|\underline{X}), c = 0, 1$, denote the posterior probability on health status. Baseline logistic regression (LR) posits a log-linear odds ratio, i.e.

$$\log \frac{P(C=1|\underline{X})}{P(C=0|\underline{X})} = \sum_{i=1}^{N} \beta_i X_i + \beta_0, \tag{1}$$

or, equivalently, that

$$P(C=1|\underline{X}) = \frac{e^{\sum_{i=1}^{N} \beta_i X_i + \beta_0}}{1 + e^{\sum_{i=1}^{N} \beta_i X_i + \beta_0}}, \tag{2}$$

where $\underline{\beta} = (\beta_0, \beta_1, \ldots, \beta_N)$ are the model parameters, estimated based on the given case-control population. An interaction between factors $X_i$ and $X_j$ (and, thus, formation of the alternative hypothesis) is introduced by adding a multiplicative term $\gamma_{ij} X_i X_j$ to the exponents in (2). The candidate interaction's significance is measured by choosing as the test statistic twice the difference between the population's data log-likelihood based on the model that includes the interaction term and the log-likelihood for the baseline model. According to Wilks' theorem (Wilks(1938)), this statistic asymptotically follows a chi-squared distribution.

There are several points supporting use of the logistic regression framework to specify the baseline (null) model. First, the range of the log-transformed odds ratio matches that of a linear combination of predictor factors, as both span the whole real line. Second, the maximum likelihood optimization problem for estimating the LR model parameters $\beta$ is convex and thus amenable to standard optimization techniques that yield globally optimal parameter estimates. Finally, there is great familiarity with LR and wide availability of LR modeling software in standard statistical packages. Accordingly, LR (Agresti(2002)) and its case-only study form (Cordell(2009)), (Vanderweele(2011)) have become de facto standards for detecting interactions in case-control studies in genetics, medicine, and in the social sciences. However, there are several limitations of this approach, outlined below.

**Theoretical Limitations of the LR Null**

In general, the "correct" null form is both unknown and domain-dependent. Thus, in choosing a model one must be guided by several desiderata: 1) theoretical plausibility – is the model derivable starting from plausible assumptions and domain knowledge ?; 2) model parsimony; 3) model *robustness* in the face of commonly occurring confounding effects; 4) experimental support/validation. In the case of LR, several considerations are not well met, as outlined below.

*LR framework is (in some instances) biologically implausible*

LR is not particularly inspired by nor derivable from an underlying conceptual model of disease risk. Indeed, it was originally proposed to model population growth (Verhulst(1838)). One feature of LR that is explicitly inconsistent with many models of disease is its symmetry or exchangeability with respect to the class label status. Specifically, when modeling the relationship between two or more risk factors and a binary disease outcome, a common conceptual model is that one may get the disease if any of the risk factors are penetrant or active, whereas being healthy requires all of the factors to be inactive. This conceptual model is inherently asymmetric with respect to the two statuses, disease and health. However, in LR, the two statuses are exchangeable and the model is thus symmetric, i.e. $\log(P_{LR}(C=1|\underline{X})/P_{LR}(C=0|\underline{X}) = \sum_{i=1}^{N} \beta_i X_i + \beta_0$, and if we swap disease and health labels, that is, $C=1 \Leftrightarrow C'=0$ and $C=0 \Leftrightarrow C'=1$, we have $\log(P_{LR}(C'=1|\underline{X})/P_{LR}(C'=0|\underline{X}) = \sum_{i=1}^{N}(-\beta_i)X_i + (-\beta_0)$. That is, the LR form is invariant to label swapping; moreover, the data likelihood, with estimated parameters plugged in, is also invariant to label swapping. That is, for LR it does not matter whether the case group is defined to consist of disease or healthy subjects. Equivalently, under LR, one does not require all the risk factors to be inactive in order to be healthy. Even without more detailed consideration, LR, as a symmetric health status model, thus appears to be in conflict with a common conceptual model of disease risk.

*Invalidity of LR in the setting of missing or surrogate causal factors*

Another important consideration, in choosing a model, concerns its "robustness" in the face of confounding effects that are common in practice. The *prevailing* scenario in inquiry involving complex diseases is that one has incomplete knowledge of the true risk factors. Such "incompleteness" may take several forms, e.g. 1) missing, i.e. unmeasured factors and 2) measured factors that are surrogates which are not perfectly correlated with the true factors. There are several reasons why some true factors may not be measured. First, for many diseases, it is quite likely there are causal agents that have not yet been discovered or even postulated as part of prevailing theories and models. Given that many environmental factors may only contribute to disease through gene-environment joint effects, which are only recently being extensively investigated, one can expect that many environmental agents may have been overlooked to date. Second, even if known, some variables are difficult and/or costly to measure, and their measurement may also be unacceptably invasive. Thus, one would expect that it is the rule, rather than the exception, that some causal factors are missing from a given case-control study design. Much of the same reasoning leads to the conclusion that, rather than the true causal agents, most studies will also involve some surrogate factors, at best (relatively) strongly correlated with the true agents. Again, this may be attributable to an imperfect working theory, and also to considerations of cost, convenience, and privacy in data measurement and collection.

We would expect that our chosen model should behave robustly in the presence of these effects in particular, that they can be accurately compensated for through suitable choice of the models parameter values. However, the LR parametric form is not invariant to these two effects and there is no way to "correct" the LR model for these potentially confounding effects in practice. As an example, suppose there are three binary causal factors ($N = 3$) and that the LR baseline joint effects model, when all three binary causal factors are observed, has parameter values: $\beta_0 = -4$, $\beta_1 = 2$, $\beta_2 = 2$, and $\beta_3 = 2$. Table 1 shows the disease probability under each of the eight combinations. Suppose now that the third risk factor $X_3$ is not observed. Further, suppose that $P[X_3 = 0] = P[X_3 = 1] = 0.5$. Then the distribution of the disease probability for this case (with $X_3$ unobserved) is obtained by averaging the left and right sub-tables of Table 1. The new distribution is shown in Table 2. Assuming that the LR model form is preserved when $X_3$ is not observed, let us denote by $\beta'_0$, $\beta'_1$, and $\beta'_2$ the parameter values for the new LR model. As there are three parameters, any three combinations from the four combinations in Table 2 suffice to calculate the parameters. We choose to compute these by excluding $(X_1 = 1, X_2 = 1)$, as follows:

$$\beta'_0 = \log\left(\frac{P(C=1|X_1=0, X_2=0)}{1 - P(C=1|X_1=0, X_2=0)}\right) = \log\left(\frac{0.0686}{1 - 0.0686}\right) = -2.6084$$

$$\beta'_1 = \log\left(\frac{P(C=1|X_1=0, X_2=1)}{1 - P(C=1|X_1=0, X_2=1)}\right) - \beta'_0 = \log\left(\frac{0.3096}{1 - 0.3096}\right) + 0.0711 = 1.8064$$

$$\beta'_2 = \log\left(\frac{P(C=1|X_1=1, X_2=0)}{1 - P(C=1|X_1=1, X_2=0)}\right) - \beta'_0 = \log\left(\frac{0.0686}{1 - 0.0686}\right) + 0.0711 = 1.8064$$

The odds for the combination $(X_1 = 1, X_2 = 1)$ based on the LR model is:

$$\frac{P(C=1|X_1=1, X_2=1)}{1 - P(C=1|X_1=1, X_2=1)} = e^{\beta'_0 + 2\beta'_1} = 2.7385. \tag{3}$$

However, from Table 2, the odds for $(X_1 = 1, X_2 = 1)$ is $\frac{0.6904}{1-0.6904} = 2.2300$. Thus, the assumption that the LR form is preserved when there are missing factors (*i.e.*, that estimating new LR parameter values only for the observed factors is equivalent to marginalizing over the missing factors) is contradicted in this example and, thus, as a general rule.

|  | X3 = 0 | |  | X3 = 1 | |
|---|---|---|---|---|---|
| P | X2 = 0 | X2 = 1 | p | X2 = 0 | X2 = 1 |
| X1 = 0 | 0.0180 | 0.1192 | X1 = 0 | 0.1192 | 0.5000 |
| X1 = 1 | 0.1192 | 0.5000 | X1 = 1 | 0.5000 | 0.8808 |

Table 1. Posterior probability of disease, with all risk factors observed, under the logistic regression model.

| p | X2 = 0 | X2 = 1 |
|---|---|---|
| X1 = 0 | 0.0686 | 0.3096 |
| X1 = 1 | 0.3096 | 0.6904 |

Table 2. Posterior probability of disease, with risk factor $X_3$ missing, under the logistic regression model.

In a similar fashion, also by counterexample, one can show that the LR form is *also* not invariant to observation of *surrogate* factors, imperfectly correlated with the true factors, rather than the true ones. That is, properly accounting for the correlation structure between measured surrogates and their "upstream" causal factors does not preserve the LR model parametric form. This is demonstrated in Appendix A. "Correcting" the LR model to account for these effects would require: 1) initially, estimating the "true" LR model (based on observation of all causal factors) and then 2) transforming this model to account for the missing (or surrogate) causal factors. In the missing factor case, step 2 would involve marginalizing out the missing factors while in the surrogate factor case, it would involve accounting for the correlation structure between surrogate and causal factors (see Appendix A). However, since in practice it is *unknown* which (and how many) causal factors may be missing (or replaced by surrogates) and since, anyway, under these two scenarios, there are no data observations for these causal factors, there is no way to implement either of these two correction steps. Thus, both missing causal and surrogate factors are *confounding* effects for LR, with no way in practice to correct the LR model to account for them. In the presence of these effects, the LR model will be biased which, as will be shown experimentally below, results in inflated type 1 error. Since the case-only LR approach (Cordell(2009)),(Vanderweele(2011)) is based on the same principle as LR, it suffers from the same deficiencies. Moreover, case-only is claimed to gain power over LR through the strong assumption that predictors are independent, which may not be satisfied in practice.

In the next section, we develop a new model, AIM, which: i) unlike LR, is inspired by an intuitive underlying disease model; ii) possesses similar model parsimony as LR (i.e., a log-additive form); iii) is however *asymmetric* with respect to diseased and healthy statuses; and iv) does possess theoretical invariance properties that make the model robust to missing and surrogate factors. Moreover, the model possesses other attractive properties (consistency under multiple disease sub-types). Finally, we give a precise, operational definition of a type of interaction commonly encountered in practice – "synergistic" – for which we mathematically prove that AIM has greater detection power than LR. In a variety of experiments involving "synergistic" interactions, using both real and simulated data sets, we demonstrate substantial gains in power of AIM over LR. Moreover, in controlled simulation experiments to investigate the effects of missing, surrogate factor, and complex disease (with subtypes) scenarios, we demonstrate that LR has inflated type 1 error, while AIMs type 1 error is resilient, in the presence of these confounding effects.

## 2  Asymmetric Independence Model (AIM)

To help develop the AIM model, we first (without any loss of generality) impart a particular interpretation to the values $\{0, 1\}$, taken on by the factor variable $X_i$ – we will construe $X_i = 1$ as meaning that the $i$-th disease factor is *active*, with $X_i = 0$ indicating the factor is inactive. We then make the following two assumptions, which together determine the form of the AIM model.

**Assumption 1:** Factors *independently* exert effects on health status ('diseased' or 'healthy'). This assumption translates mathematically as follows. We define a *latent* "local" binary disease status random variable $C_i \in \{0,1\}$ coupled to each factor $X_i$, i.e. with the $C_i$ assumed statistically independent of each other given the status of $X_i$. Supposing we have $N$ factors, this assumption is mathematically equivalent to $P[C_1, \ldots, C_N | X_1, \ldots, X_N] = \prod_{i=1}^{N} P[C_i | X_i]$. $C_i$ is defined such that $P[C_i = 1 | X_i = 0] = 0$, i.e. the factor being active is required for the *local* status to be 'diseased'. However, the factor's active status does not deterministically cause the local status to be 'diseased' – it gives propensity for the local status to be diseased, based on the conditional probability $\phi_i \equiv P[C_i = 1 | X_i = 1]$. How the individual local disease statuses jointly contribute to/determine the *global* disease status $C$ is specified next.

**Assumption 2:** an individual is healthy only if every factor that is active does not cause its local status to be 'diseased', i.e. $C = 0 \Leftrightarrow (C_0 = 0) \cap (C_1 = 0) \cap (C_2 = 0) \cdots (C_N = 0)$. Here, $C_0$ is a "background" factor accounting for sporadic disease occurrence, with probability $\phi_0 \equiv P[C_0 = 1]$.

**Stochastic Data Generation**

The above two assumptions fully specify a stochastic data generation mechanism for the disease status $C$ given an observed factor vector $\underline{X} = \underline{x}$, as follows: 1)Independently randomly generate $C_i$ given $x_i$, $i = 1, \ldots, N$, according to the probability model defined under Assumption 1, and randomly generate $C_0$ according to the pmf $\{1 - \phi_0, \phi_0\}$. 2)Assign $C = 0$ if $(C_0 = 0) \cap (C_1 = 0) \cap (C_2 = 0) \cdots (C_N = 0)$. Otherwise, assign $C = 1$.

**Posterior Probability Model**

The posterior probability of 'healthy' status, given a factor vector $\underline{x}$, consistent with the above stochastic data generation, is:

$$P(C = 0 | \underline{x}) = (1 - \phi_0) \prod_{i=1}^{N} (1 - \phi_i)^{x_i}. \tag{4}$$

Taking the logarithm, we obtain

$$\log P_{\text{AIM}}(C = 0 | \underline{x}) = \log(1 - \phi_0) + \sum_{i=1}^{N} x_i \log(1 - \phi_i) = \beta_0 + \sum_{i=1}^{N} \beta_i x_i, \tag{5}$$

where $\beta_0 \equiv \log(1 - \phi_0)$ and $\beta_i \equiv \log(1 - \phi_i)$. That is, whereas in logistic regression the log-odds are linear in the factors, in AIM, the log-probability of health is linear. Exponentiating, we obtain the AIM posterior probability of healthy status as:

$$P_{\text{AIM}}(C = 0 | \underline{x}) = e^{\beta_0 + \sum_{i=1}^{N} \beta_i x_i} \tag{6}$$

and, thus, with the posterior probability of disease:

$$P_{\text{AIM}}(C=1|\underline{x}) = 1 - P(C=0|\underline{x}) = 1 - e^{\beta_0 + \sum_{i=1}^{N} \beta_i x_i}. \tag{7}$$

Note that in general we must have $0 \leq P_{\text{AIM}}(C=0|\underline{x}) \leq 1 \forall \underline{x}$. To ensure this, we would require $\beta_0 + \sum_{i=1}^{N} \beta_i x_i \leq 0, \forall \underline{x} \in \{0,1\}^N$, with number of constraints exponential in $N$. However, since the model is being estimated from a finite population $\mathcal{X} = \{\underline{x}^{(1)}, \ldots, \underline{x}^{(T)}\}$ and used solely for hypothesis testing on this population, the model need only meet these constraints on the given population, i.e. we really only require $\beta_0 + \sum_{i=1}^{N} \beta_i x_i^{(n)} \leq 0, \forall \underline{x}^{(n)} \in \mathcal{X}$.

Note also that while it is possible to do so, we do *not* constrain the $\beta_i$ to be negative (consistent with $\beta_i$ being the log of a probability). In this way, the estimated model may contain factors which attenuate disease risk ($e^{\beta_i}, \beta_i > 0$) as well as those which amplify this risk ($e^{\beta_j}, \beta_j < 0$).

**AIM Model Learning and Hypothesis Testing**

In Appendix B, we show that AIM's maximum likelihood (ML) objective function is concave in the parameters $\underline{\beta}$, with the constrained ML learning problem (concave objective, linear constraints) thus a convex optimization problem (Boyd(2003)), amenable to finding the global maximum. We propose a hybrid Newton-Barrier function algorithm for its solution (Appendix C). Statistical interaction detection for AIM, like LR, is based on a log-likelihood ratio statistic which, under the null, according to Wilk's Theorem (Wilks(1938)), is asymptotically chi-squared.

**Asymmetry of AIM**

As discussed earlier, since mechanisms of being healthy and diseased are different, a model for disease risk should not bear the same parametric form under both statuses. Recall that LR is a symmetric model, violating this basic biological constraint. However, AIM is asymmetric, as seen by noting that while its log-probability of being healthy is a linear function of the factors, the log-probability of being diseased is clearly nonlinear. Although AIM and LR are thus very different models, there is an extreme setting under which the two models will coincide. In particular, if $P(C=1|\underline{x}) \to 1$ and, thus, $\log P(C=1|\underline{x}) \to 0$, we have $\log(P(C=1|\underline{x})/(1-P(C=1|\underline{x}))) \sim -\log(1-P(C=1|\underline{x}))$. In this case, the two models converge to a common model. However, in practice, the fraction of cases will typically be smaller than the fraction of controls (as cases are often harder to obtain than controls). In this context, as well as for the other common scenario of a balanced case-control population, the LR and AIM models are quite different.

**Link to Well-Established Biological Theories of Disease**

In addition to its asymmetry, AIM is supported by several well-accepted biological models, including the heterogeneity theory (McClellan(2010)) and the two-hits theory of cancer (Knudson(2001)). The former states that any one of many different mutations in any one of many different genes leads to related phenotypes. Take hearing loss (McClellan(2010)) as an example. Here, the responsible genes encode proteins involved in a wide variety of processes in the inner ear, including development and maintenance of cytoskeletal structures, myosin motors, gap junction transport and signaling, ion channels, and transcriptional regulators. Consistent with Assumption 2) in our model, if any of these processes fails, the person will lose hearing. It is also reasonable to have Assumption 1) that factors independently exert effects, due to separation of the functional modules. The two-hits theory of cancer makes Assumption 1) even more compelling. When one individual possesses some disease-risk factors (often germline mutations), this is called the first hit. Disease will not develop until the second hit – random somatic mutation – occurs. Hence, each disease-risk factor exerts its effect through random somatic mutation. Moreover, in general, random somatic mutations for different genes are expected to be independent.

While we have argued that AIM is more biologically plausible than LR, we believe the most compelling support for AIM comes from the *invariance* of this model, unlike LR, in the presence of unmeasured, surrogate factor, and disease heterogeneity (subtype) confounding effects.

**Model Consistency in the Presence of both Unmeasured and Surrogate Risk Factors**

The pervasiveness of missing and surrogate factors (as previously discussed) raises fundamental questions for models of joint baseline effects: 1) Is the model consistent (i.e. is the model's form preserved) when there are missing and surrogate risk factors? 2) If the model form is not preserved, what are the performance implications? We will resolve 2) empirically through our experiments. To resolve 1), we have the following theorems, proved in Appendices D-G.

**Theorem 1: Assuming statistically independent factors, AIM is a consistent model when there are unmeasured (missing) causal factors, while LR is not. Moreover, under the AIM model, the parameter values themselves ($\{\beta_i\}$) for the observed factors do not change, in the presence of missing factors.**

**Theorem 2: Assume that some causal factors are not measured, but surrogate factors, correlated with these true factors are instead measured. Assume the following statistical dependency structure: a causal factor $X_i$ is conditionally independent of all other factors (either the true factors or their surrogates), given the causal factor's surrogate,**

$X_i^0$. Also assume that the disease status is conditionally independent of a surrogate factor, given the true factor. Then, the AIM model is consistent under the surrogate factors scenario, while the LR model is not. Moreover, under the AIM model, the parameter values themselves ($\{\beta_i\}$) for the observed true factors do not change, in the presence of surrogate factors.

*Comments:*

1) For LR, the theorems are proved by counterexample (as already shown in the previous section for the unmeasured factors scenario). 2) In the unmeasured case, the new model form is obtained by marginalizing (integrating out) unmeasured factors. Marginalization of AIM leads to the same mathematical model form, while this is not true for LR. To understand why the AIM form is preserved under unmeasured factors, note that $P_{\text{AIM}}(C = 0|\underline{x}) = e^{\beta_0} \prod_{i=1}^{N} e^{\beta_i x_i}$. Thus, when a factor is not measured, it is essentially omitted from the product – this effects marginalization, and preserves AIM's log-additive form on the remaining factors (A formal proof is given in Appendix D.). The proof for consistency under surrogates is given in Appendix E. The practical implication of these theorems, demonstrated in the Experimental results section, is that LR has inflated type 1 error under these scenarios, while AIM does not. Moreover, AIM has greater detection power than LR under these scenarios; 3) The rigorous proofs of Theorems 1 and 2 require the assumption of independence between factors. This assumption may not hold for some applications. We use simulations to investigate the implications of the violation of this assumption. As shown in section 3.1.5, we do not observe any inflation in type I error rate when there are missing factors, suggesting Theorem 1 remains practically valid. On the other hand, we do see that Theorem 2 cannot be true when the independence assumption is violated. However, the effect is not detectable when the correlation is moderate, and the effect is still small when the correlation is very strong.

**Model Consistency Under Disease Heterogeneity**

In addition to its consistency under these two confounding scenarios, AIM is *also* a consistent model, and LR an inconsistent one, with respect to yet another confounding source – disease *heterogeneity*. Specifically, suppose that there are several disease subtypes $D_i \in \{0, 1\}, i = 1, \ldots, K$, where $D_i = 1$ means the $i$-th subtype is present in an individual. Likewise, the heterogeneous disease is present, i.e. $C = 1$, if and only if at least one disease subtype is present, i.e. if and only if $(D_1 = 1) \bigcup (D_2 = 1) \cdots \bigcup (D_K = 1)$. If the different subtypes are known *and* if the cases in the population were ground-truth labeled by subtype, one could estimate a separate case-control posterior model quantifying the baseline risk for each disease subtype $P(D_i|\underline{x})$. However, in practice, this is unrealistic – a complex disease may decompose as subtypes, but these will typically be *latent*, with explicit knowledge only of whether the heterogeneous disease is present, not which subtype. Regardless of whether subtypes are explicitly known or not, there is a posterior for each disease subtype $P(D_i|\underline{x}), i = 1, \ldots, K$. Moreover, a model for the complex disease status is the posterior $P(C = 1|\underline{x}) = P((D_1 = 1) \bigcup (D_2 = 1) \cdots \bigcup (D_K = 1)|\underline{x})$. If the individual subtype models are AIM models, and if disease subtypes are conditionally independent given the observed factors, then one can show that the complex disease model is *also* an AIM model, i.e. the AIM parametric form is invariant to disease heterogeneity. On the other hand, this is again not true for LR. Specifically, we have:

**Theorem 3: Suppose that a complex disease contains multiple subtypes, which are assumed to be conditionally independent given the observed factors. Then, the AIM model form is invariant to disease heterogeneity, i.e. $P((D_1 = 1) \bigcup (D_2 = 1) \cdots \bigcup (D_K = 1)|\underline{x}) = P_{\text{AIM}}(C = 1|\underline{x})$, where, in particular, the weight $\beta_i$ on an individual factor $X_i$ in the heterogeneous AIM model is *additive* over the weights on this factor for each of the disease subtype models. On the other hand, the LR form is not invariant to disease heterogeneity.**

The proof of this Theorem is given in Appendix F. An important implication of this theorem is the following: to do inference on the heterogeneous disease using the AIM model, one need not have *any* prior knowledge of how many (and whether in fact) multiple disease subtypes exist for the given disease domain. The AIM modeling approach is naturally accommodating of however many disease subtypes that may be present (through the additive weight mechanism).

**Theoretical Characterization of Detection Power for AIM and LR**

Generally speaking, for a two-sided hypothesis testing problem it is difficult to draw a uniform conclusion on the power comparison between two competing models. In fact, in general a "no free lunch theorem" should apply, with no model/method uniformly dominating another. Thus, it is very useful to identify the conditions or assumptions under which one model is theoretically guaranteed to outperform another. Such results can inform when it is most suitable in practice to apply one model, rather than another. We have identified conditions under which AIM is guaranteed to perform better than LR. Strongly supporting the usefulness of AIM, these conditions correspond to the most common scenarios encountered in real applications. Consider two types of interactions: (1) synergistic and (2) antagonistic. A synergistic interaction means that the true effect associated with the joint occurrence of two risk factors is greater than a baseline model's (without interaction) joint effect (Phillips(2008)). On the other hand, if the true joint effect is smaller than a baseline model's (without interaction) joint effect, we call it "antagonistic". Most interactions found in practice are synergistic (Phillips(2008)). Theorem 4 below, based on a precise, meaningful, and *operational* definition of synergistic interactions, shows that AIM has better power to detect synergistic interactions than LR. A corollary can also be derived stating that LR is guaranteed better power than AIM for antagonistic interactions. However, even for antagonistic interactions we argue against the use of LR because of its degraded performance when there are missing and surrogate factors and/or disease subtypes.

**Theorem 4: Given two binary variables $X_1$ and $X_2$, let $p_{00} \equiv P(C = 1|X_1 = 0, X_2 = 0)$ and similarly define $p_{01}$, $p_{10}$, and $p_{11}$. Assume $p_{01} \geq p_{00}$ and $p_{10} \geq p_{00}$. Denote $p'_{11}$ the predicted value from the LR model whose three parameters are determined *solely* by the true posterior probabilities $p_{00}$, $p_{10}$, and $p_{01}$ (This model is precisely defined in Appendix G). We define a synergistic interaction as one satisfying $p_{11} \geq p'_{11}$. Under the above assumptions, we then have the following result: for synergistic interactions, AIM gives a greater difference between its interaction model and baseline model log-likelihoods than that for LR. Hence, AIM generates a strictly smaller p-value than LR and hence has better power to detect interaction effects than LR. Likewise, if $p_{11} < p'_{11}$, *i.e.* an antagonistic interaction, then LR generates a smaller p-value than AIM.**

*Proof:* See Appendix G.

### Applicability for Non-binary Factors

Both in the above derivation of AIM and in developing its theoretical properties, we assumed that factors are binary. All of the above results can be straightforwardly extended for the case where factors are non-binary but categorical. In particular, a nonbinary categorical variable $X$ with cardinality $L$ can be recoded as a vector of $L$ binary factors $\in \{0, 1\}$, with only one of these factors "on" to specify a value for $X$. The AIM model can also be *applied* when the variables $X_i$ are quantitative (or ordinal). However, the AIM model form is not logically derivable in the same way as given above for binary factors. Moreover, while AIM's invariance properties hold for nonbinary and quantitative factors (since we do not assume factors are binary in the proofs of theorems 1,2, and 3 given in Appendices D,E, and F, respectively), the rigorous proof of Theorem 4 on detection power relies on the assumption of binary factors.

# 3 Simulation Study

Our simulation study evaluates the type 1 error and detection power of AIM and logistic regression in a controlled setting, under varying parameter settings which characterize the population being studied, and under the three confounding scenarios prominently identified in this paper – missing factors, surrogate factors, and disease subtypes. The goal is to understand the performance effects of different parameter settings and of these scenarios on both models. These results are next given.

## 3.1 Evaluation of Type I Errors

Figure 1 shows the Q-Q plot for AIM with 1000 trials. All trials were simulated by randomly drawing samples from an AIM model with two independent factors and parameters $\beta_0 = -0.337$, $\beta_1 = \beta_2 = -0.336$. The parameters were chosen so that the case-control ratio is around 1 and the marginal effect size for each factor is an odds ratio of 2. We further assumed all 4 factor combinations are equally likely. In each experiment, 4000 samples were generated. We can see that the Q-Q plot closely aligns with the diagonal line, indicating an accurate assessment of significance for AIM. We also generated Q-Q plots for dependent factors and unbalanced factor combinations. Indeed, we have simulated the dependent factors with correlation coefficient as large as 0.9. We also made one of the four factor combinations as small as 1 %. We do not observe any obvious deviation from the diagonal line for these Q-Q plots.

### 3.1.1 Varying case fractions

Figure 2 shows the empirical type I error (evaluated when the null hypothesis of no interaction is valid) at significance level 0.05. The gray region is the 95% confidence interval. We assessed the influence of the case fraction on the empirical type I error. Each estimate is based on 1000 tests, with the empirical type I error calculated as the ratio between the number of tests that have p-value smaller than 0.05 and the total number of tests. The AIM model used to generate the data here is: $\log(1 - P(C = 1|\underline{x})) = a(-0.337 - 0.336x_1 - 0.336x_2)$, with $a$ varied to sweep the range of case fractions from 0.05 to 0.95. The factor distribution is the same as assumed for Figure 1 experiments. We can see that for all scenarios the empirical type I error closely approximates the expected type I error.

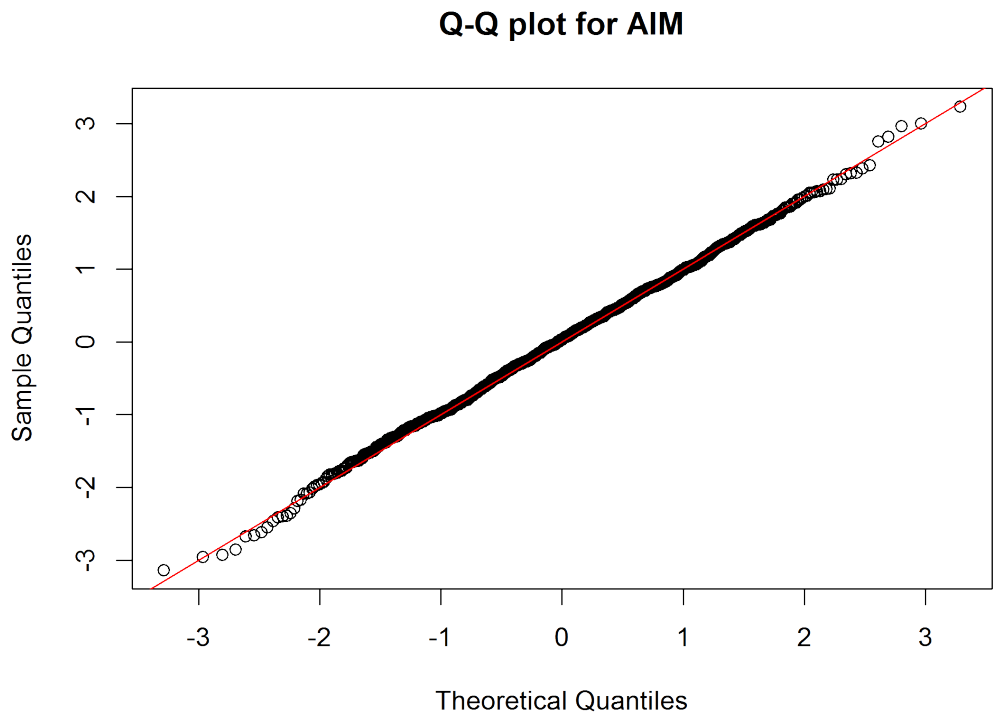

Figure 1: Q-Q plot for AIM.

### 3.1.2 Missing factors

Figures 3a, 3b, and 3c show the empirical type I error rate at significance level 0.05 for LR and AIM. For each comparison, the left (dark) bar is LR's measure and the right (light) bar is AIM's. This is the convention used in all our type 1 error figures. For AIM experiments, the data was generated according to the baseline AIM model, and for LR experiments, the data was generated according to the baseline LR model, with the ground-truth AIM and LR model parameter values chosen so that the marginal effects of each of the factors was the same for the two (AIM and LR) baseline models. We simulated in total 18 scenarios with different case fractions and number of missing factors. To get reliable estimates of type I error, 10,000 datasets were simulated for each scenario. Both the point estimate and the 95% confidence interval are shown in the figure. All the scenarios are designed to have marginal effects with an odds ratio of 2 for the observable factors.

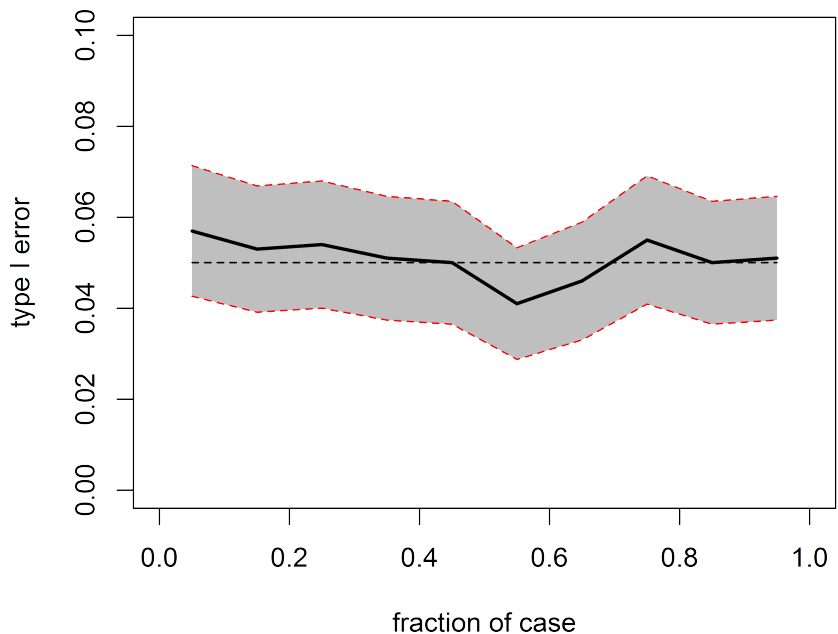

Figure 2: Empirical type I error of AIM with varying case fraction.

In Figure 3a, we simulated one missing factor with effect size of 15. In Figure 3b, we simulated 10 missing factors with comparable effect sizes as for the observable factors. In Figure 3c we simulated 100 missing factors with effect sizes of 1.1. In all scenarios, the empirical type I error rates for AIM match the theoretical value (0.05) very well. The inflation of type I error rate for LR has multiple causes. However, generally speaking, the fewer number and the larger effect sizes of missing factors result in larger inflation. Also seen from the figure, larger inflation occurs when the case-control ratio deviates from balanced (0.5). Interestingly, we observed no inflation for LR when the case fraction is 0.5.

### 3.1.3 Surrogate markers

Figures 4a and 4b compare the empirical type I error rates for both LR and AIM when the observed markers are surrogate instead of causal. Each empirical type I error rate is estimated based on 10000 experiments. The dashed lines indicate the expected type I error rate and the 95% confidence intervals for each estimate are marked by the corresponding error bars. In Figure 4a, the effect size for the observable surrogate markers is small and approximately 1.5 in terms of odds ratio. In Figure 4b, the effect size is around 5. The empirical type I error for AIM is close to the expected in all cases. The accuracy of type I error rate for LR is dependent on two factors – the effect size and the degree of correlation between a surrogate and its associated causal factor ($r^2$). Larger effect size and weaker correlation generally imply larger deviation from the expected value.

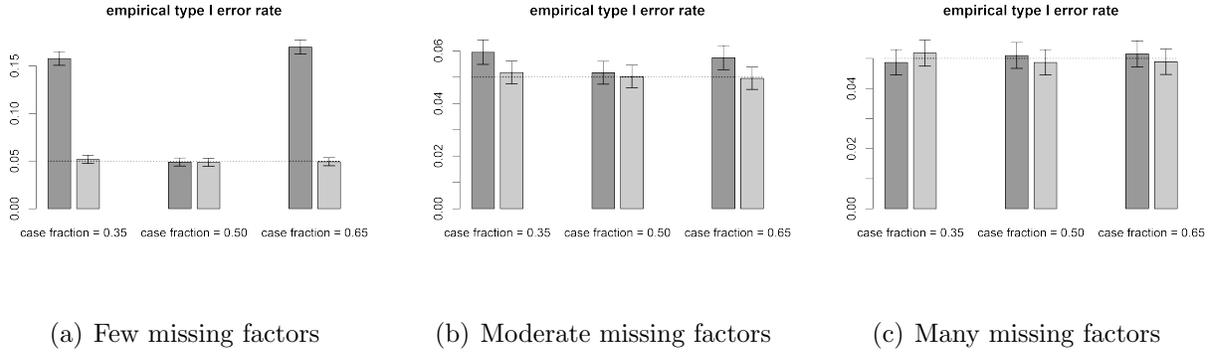

(a) Few missing factors     (b) Moderate missing factors     (c) Many missing factors

Figure 3: Empirical type I error rate at significance level 0.05 for LR (dark grey) and AIM (light grey) when there are a) a few missing factors with large effect size; b) a moderate number of missing factors with moderate effect size; and c) when there are a lot of missing factors with small effect size.

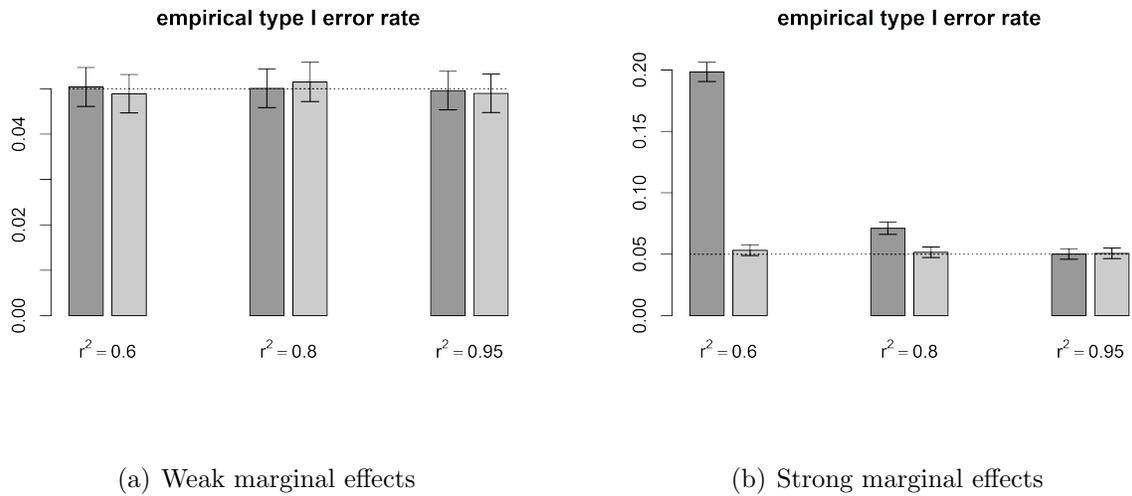

(a) Weak marginal effects          (b) Strong marginal effects

Figure 4: Empirical type I error rate at significance level 0.05 for LR (dark grey) and AIM (light grey) when the surrogate markers have a) weak marginal effects and b) strong marginal effects; $r^2$ is the correlation between a surrogate factor and its associated causal factor.

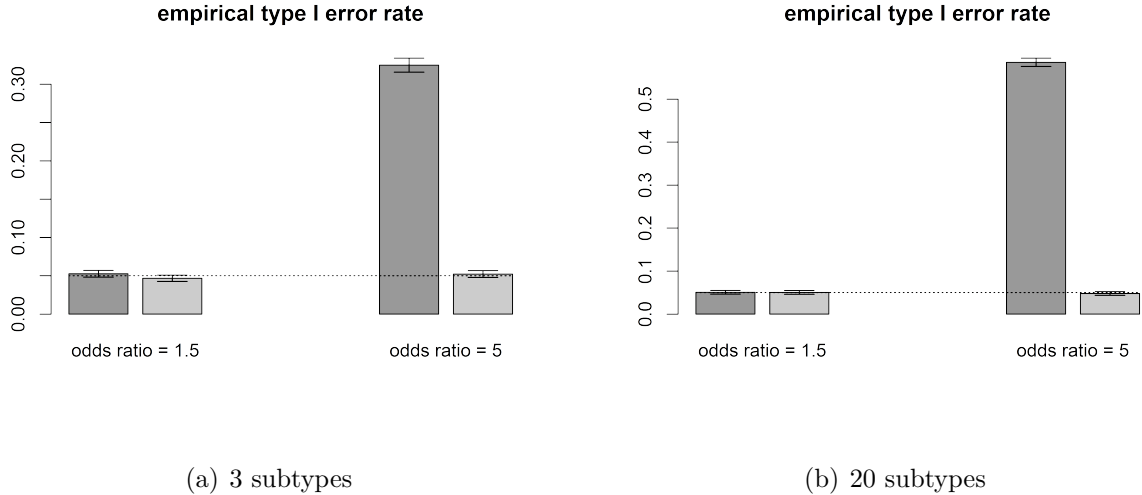

(a) 3 subtypes

(b) 20 subtypes

Figure 5: Empirical type I error rate at significance level 0.05 for LR (dark grey) and AIM (light grey) when there are a) 3 subtypes and b) 20 subtypes. Odds ratio refers to the marginal effect.

### 3.1.4 Subtypes

Figures 5a and 5b show the empirical type I errors when there are subtypes. We simulated four scenarios with different effect sizes and number of subtypes. Here we assume both risk factors have the same marginal effect size. When the effect size is large, the overall distribution deviates from the null logistic regression model significantly, though each subtype follows the logistic regression model. Each subtype independently generates the status of case or control, with the overall status a 'case' if either subtype status is a 'case'. We also notice that when the effect size is weak (odds ratio = 1.5), the effect of the subtypes is negligible.

### 3.1.5 Impact of the violation of the independence assumption in Theorems 1 and 2

The simulations in Subsections 3.1.2 and 3.1.3 confirm Theorems 1 and 2 under independence between the observable factors (when missing factors exist) or the causal factors (when surrogate factors are observed). These assumptions are non-trivial. Here, we investigate numerically the scenarios where the independence assumptions are violated. Our experiments show different effects of the violation of the assumption for the two theorems. Though we cannot prove it theoretically, it seems that Theorem 1 still practically holds even when the observed factors are strongly correlated. On the other hand, we do observe that the violation of the independence assumption leads to a systematic, though often small, breakdown of Theorem 2. Note that Theorem 3 does not involve the independence assumption and is not investigated here.

To test the impact of the violation of the independence assumption when missing factors exist, we tried various settings as in section 3.1.2. We do not see obvious inflation of the type I error for AIM. Based on our experience with LR, we hypothesized that the setting of a few but large-effect missing factors is the one that is most likely to demonstrate the impact. The left subfigure in Figure 6 shows the empirical type I error rate under this setting. Using an interval of 0.1, we surveyed the correlation with coefficients ranging from -0.9 to 0.9. For all 19 sets of experiments, all but r=0.2 have the expected Type I error rate falling in the 95% confidence interval. However, the only exception should also be expected due to the effect of multiple tests. In fact, we fixed r=0.2 and conducted the experiment again. This time, the expected type 1 error rate is within the 95% confidence interval. We tried multiple settings as in section 3.1.3 to test the impact of the violation on the type I error rate when surrogate factors are measured. Although for the majority of the settings the effect is not detectable, a small fraction of settings consistently show inflated type I error. The right figure of Figure 6 illustrates the effect under a typical scenario. The correlation is 0.8 between the causal and surrogate factors. The inflation becomes obvious when the correlation between the two causal factors is very strong. We do not see observable inflation for moderate correlations. To further test whether the inflation is due to some random effect, we fixed the correlation at 0.9 and increased the number of simulations to 100,000 to narrow the confidence interval of the estimated type I error rate. We got a point estimate as 0.0615 and the 95% confidence interval is [0.0599, 0.0630].

## 3.2 Power comparison on synthetic datasets

In this section we report results on various synthetic datasets to assess the performance of AIM in detecting true interaction effects. A comprehensive set of scenarios were simulated to evaluate how the power of AIM is affected by sample size, effect size, case-control ratio, risk factor allele frequency, p-value threshold, main effects, correlation between risk factors, missing factors, surrogate factors, and disease subtypes. For every experiment, the performance of LR was also evaluated, with special attention paid to the different trends observed for AIM and LR. In all of the reported experiments, power was empirically estimated based on 1000 random simulations. When we investigated the effect of one parameter, we fixed all other parameters. The default p-value threshold (alpha value) was set as 0.05. By default, we assumed risk factors are independent. In most of the experiments, the interaction models were based on a logistic regression model with non-zero interaction terms, as follows: $\log(\frac{p}{1-p}) = \beta_0 + \beta_1 x_1 + \beta_2 x_2 + \beta_3 x_1 x_2$. Here, both risk factors are binary. The odds ratio is used to represent the effect size. Thus, $e^{\beta_1}$ and $e^{\beta_2}$ are the two main effect sizes and $e^{\beta_3}$ the interaction effect size. By default, we set the main effect size for both risk factors to 1.5. The interaction effect size was also set to 1.5. The risk allele frequency, that is, the frequency of $x_i = 1, i = 1, 2$ was set to 50%; $\beta_0$ was adjusted so that the case fraction was around 50%. We had two different default settings for the sample size. When the experiment assesses the impact of interaction effect size, the sample size was set to 1000; otherwise the sample size was set to 2000. The smaller sample size was designed to survey the impact of a larger range of interaction effect sizes. The above-described interaction model was also extended to include high-order interactions and more complex interaction forms. We took advantage of existing interaction models (Li(2011)) and applied both AIM and LR to them. Specifically, five more interaction models were tested, spanning from 2-way to 5-way interactions and involving ternary variables. In the following subsections we give detailed discussion on the results for each set of experiments; finally we summarize our overall conclusions.

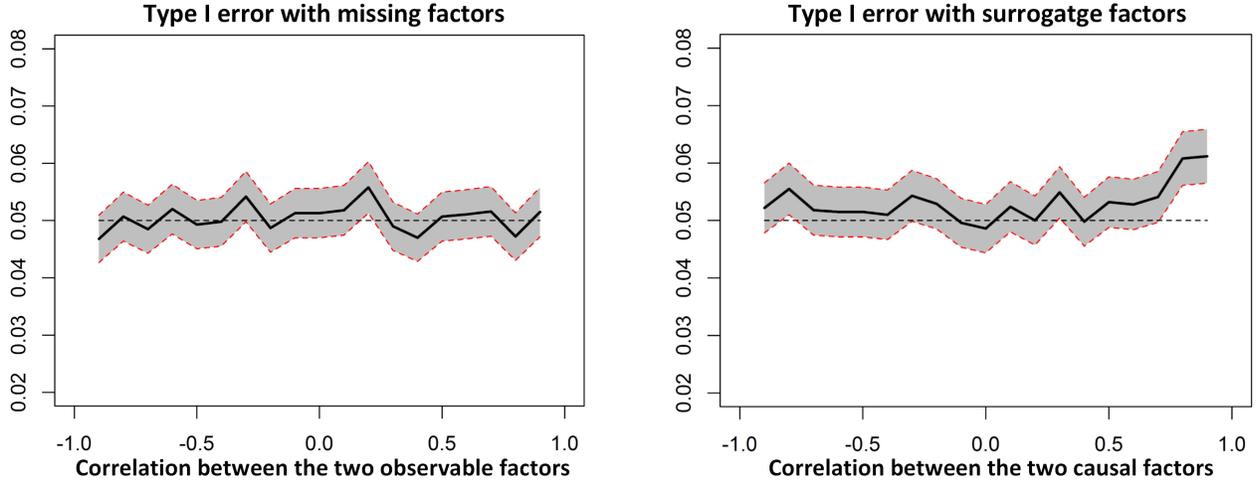

Figure 6: Empirical type I error rate at significance level 0.05 for AIM when the independence assumptions in Theorems 1 and 2 are violated. The left and right figures demonstrate the scenarios for Theorem 1 (missing factors) and Theorem 2 (surrogate factors), respectively. Red-dashed lines indicate the 95 % confidence interval.

### 3.2.1 Impact of sample size and effect size

Figures 7a, b, and c show how power is affected by sample size with the interaction effect size fixed at 1.1, 1.5, and 3, respectively. Figure 8 shows how power is affected by effect size with the sample size fixed at 1000. As expected, for both methods the power increases from 0 to 1 when either the sample size or the effect size is increased. We can also see that the effect size has more dramatic impact on power than the sample size. For AIM, 1600 samples are needed to achieve 80% power when the effect size is 1.5, compared to 280 samples when the effect size is doubled to 3.0. Under all scenarios AIM is always better powered than LR. However, the difference in power is dependent on the effect size. For example, in Figure 7a, to achieve 80% power, 13,000 and 57,000 samples are needed for AIM and LR, respectively. In Figure 7b, when the effect size is 1.5, to achieve 80% power 1,600 and 3,200 samples are needed for AIM and LR, respectively. Generally speaking, the relative gain of AIM over LR is larger for smaller effect sizes and for (relatively) small sample sizes at a given effect size, as seen in Figures 7a,b,c. This is encouraging because one often faces problems with small effect size (and limited sample size) in real applications. However, we can also observe that the maximum gain in power of AIM over LR over the range of sample sizes (as well as the area between the two power curves) decreases as the effect size increases.

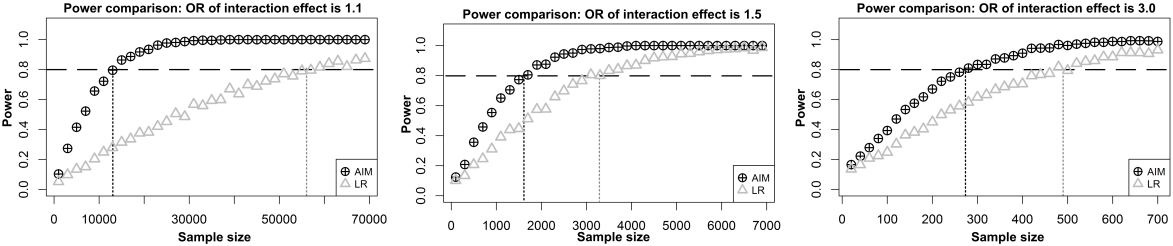

(a) Small interaction effect size  (b) Moderate interaction effect size  (c) Large interaction effect size

Figure 7: Power vs sample size when the interaction effect size is fixed at a) an odds ratio of 1.1; b) an odds ratio of 1.5; c) an odds ratio of 3. The case fraction was 50% and the main effect size was 1.5 for both risk factors.

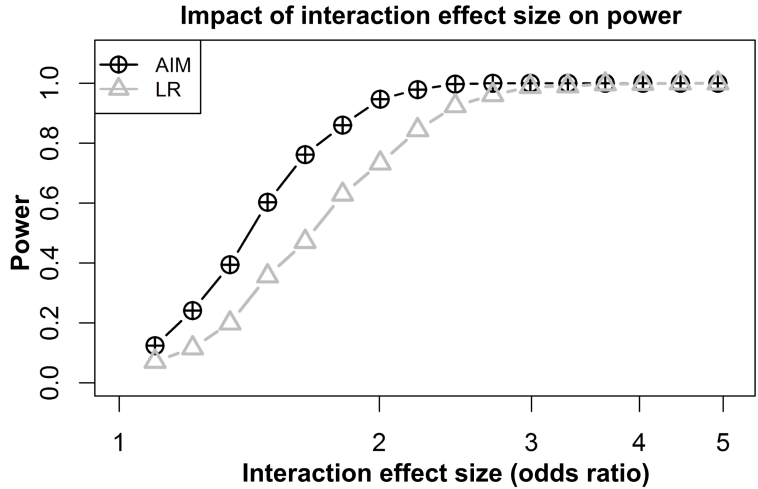

Figure 8: Power versus interaction effect size. The sample size is fixed at 1000. The main effect sizes for both risk factors were set at the odds ratio of 1.5. The case fraction is fixed at 50%.

### 3.2.2 Impact of case-control ratio

Figure 9 illustrates how the case-control ratio influences the power and how the two methods differ. We surveyed the fraction of cases from the very low end (0.1) to the very high end (0.9). It should be expected that power will not achieve its maximum at either end. Indeed, when the cohort is composed of all cases or controls, there is no way to evaluate the difference between cases and controls. Thus, the maximum power should be achieved at some intermediate value. Recalling that logistic regression is a symmetric approach with respect to case/control status, it is reassuring to see that logistic regression gets maximum power at a case fraction of 0.5. A striking difference here is that AIM achieves its maximum power when the fraction of cases is around 0.3. We can also observe that the gain of AIM over LR is larger when the fraction of cases is smaller. This phenomenon is rooted in the defining formulas for AIM and LR. Comparing the AIM form $\log(1-p)$ to the LR form $\log(p) - \log(1-p)$, where $p$ is the probability of a case, we can see that, neglecting the sign, the two forms get close when $p$ approaches 1.

### 3.2.3 Impact of risk allele frequency

Figure 10 shows how the risk allele frequency impacts power. The allele frequencies were simulated to range from 0.1 to 0.9. The power is decreased for both AIM and LR when the risk allele frequency is either extremely low or high. The power for LR is symmetric with respect to the risk allele frequency, achieving its maximum at 0.5. The power curve for AIM is skewed to the higher frequencies – when the risk allele frequency is 0.1, AIM has power of 0.4; when the risk allele frequency is 0.9, the power is 0.55. Considering that we used 1000 simulations to calculate power, this difference is highly unlikely to be due to random fluctuations. We do not have an analytical characterization of how the risk allele frequency asymmetrically affects AIM's power, with the higher allele frequencies more power-favorable. However, we believe this is again a consequence of the asymmetry of AIM with respect to case/control status.

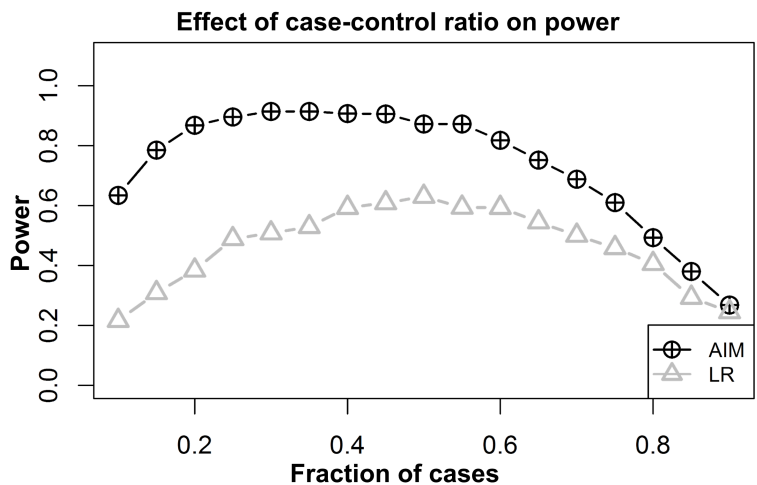

Figure 9: Power versus case-control ratio. The fraction of cases is varied by adjusting the baseline parameter $\beta_0$ in the logistic regression model possessing an interaction term. The sample size is 2000 and the interaction effect size is 1.5. The main effect sizes for both risk factors are 1.5.

### 3.2.4 Impact of main effect size

Figure 11 demonstrates how the main effect size affects the power to detect an interaction when the interaction effect size is fixed. We observe that AIM and LR follow very different trends. LR's power slightly decreases as the main effect size increases. The power of LR is 0.352 at a main effect size of 1.1 and reduces to 0.279 at a main effect size of 5.0. This decline in power is even more apparent for larger main effect sizes. Although not shown in Figure 11, we observed that LR's power falls to 0.16 for a main effect size of 20. These results are not surprising – with the interaction effect size fixed, the increase in main effect size increases the variance of the estimate of interaction effect size. On the other hand, AIM's power increases as the main effect size is increased. This can be understood by recognizing i) that the interaction effect is the difference between the true effect and the one predicted by the null hypothesis; and ii) that AIM and LR posit very different null hypotheses. In particular, in this experiment the data were generated based on an LR model, with the interaction effect size fixed while varying the main effect size. A fixed interaction effect for an LR model will almost assuredly give a *variable* interaction effect size for the AIM model, as the main effect size is varied.

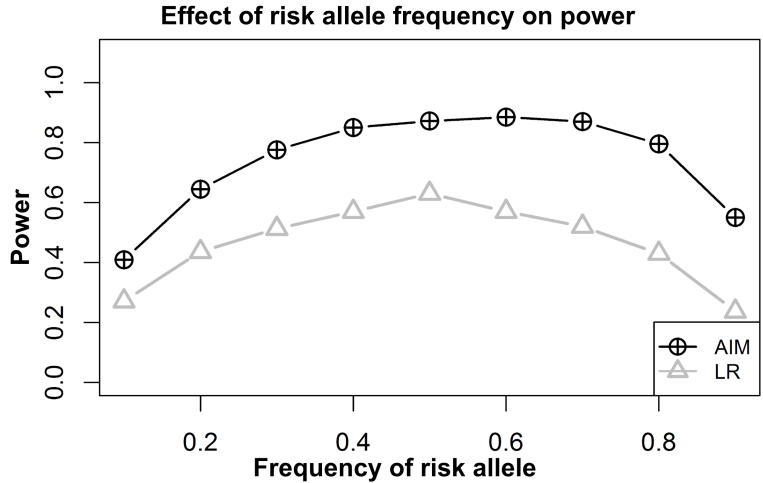

Figure 10: Power versus frequency of risk allele. The sample size is 2000. The main effect sizes for both risk factors are set as 1.5 and the interaction effect size is 1.5. In the experiments one risk factor has the allele frequency changing while the other risk factor's allele frequency is fixed. The case fraction is fixed at 50%.

### 3.2.5 Impact of p-value threshold

Figure 12 shows how the sample size needed for 80% power is dependent on the p-value threshold of significance. We varied the p-value threshold from 0.05 to 5e-8. It appears that the sample size is linearly proportional to the log-transformation of the p-value threshold, for both AIM and LR. However, AIM's slope is smaller than that of LR. When the p-value threshold is 0.05, 1600 and 3200 samples are needed for AIM and LR, respectively, *i.e.*, twice AIM's number of samples are needed for LR. When the p-value threshold is 5e-8, the ratio is 2.17 (computed as 165,000 divided by 76,000).

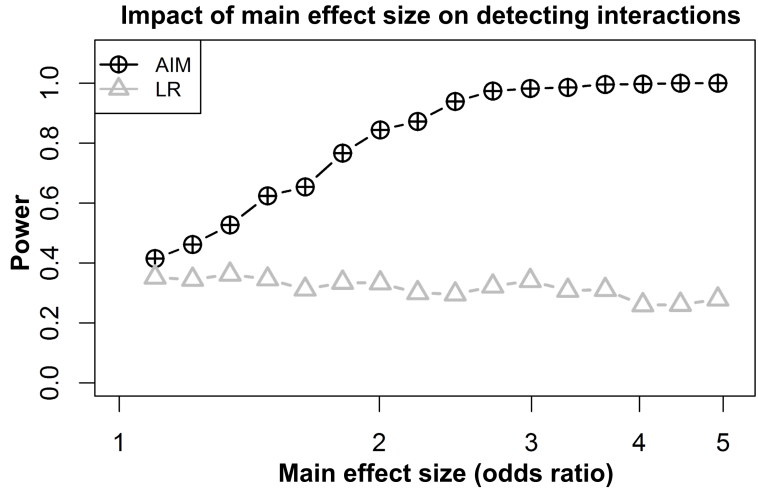

Figure 11: Power vs main effect size. The sample size is 1000 and the interaction effect size is fixed at 1.5. The case fraction is fixed at 50%.

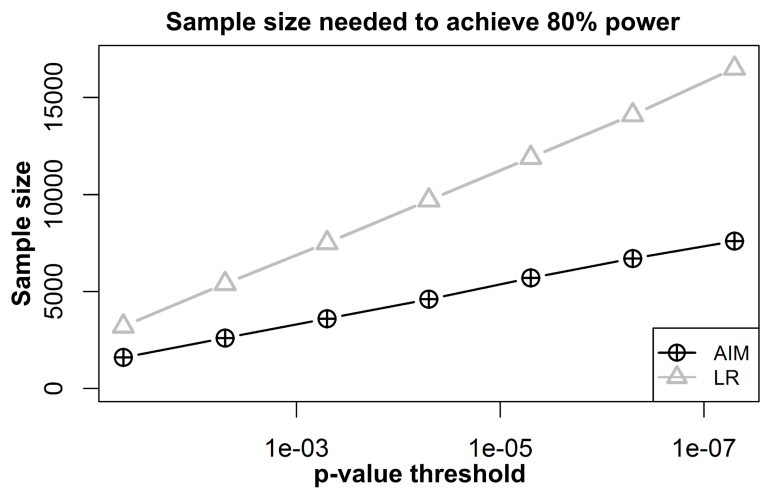

Figure 12: Sample size versus p-value threshold. The main effect size is 1.5. The interaction effect size is 1.5. The case fraction is 50%.

### 3.2.6 Impact of correlation between risk factors

Figure 13 illustrates how the correlation between the two risk factors affects power. High absolute correlation significantly reduces power. For example, when the correlation is -0.95, the power for AIM reduces from 0.902 to 0.282, while the power for LR reduces from 0.626 to 0.153. The maximum power is achieved for both AIM and LR when the two risk factors are independent. It is worth noting that unlike for the case-control ratio and allele frequency, AIM's power, similar to LR's, exhibits symmetric dependence on the correlation between the two factors.

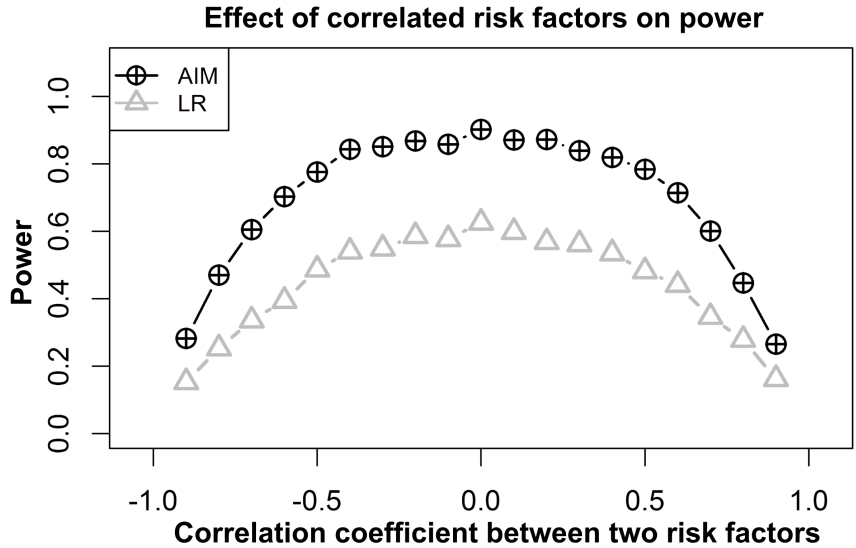

Figure 13: Power to detect an interaction versus correlation between the risk factors for AIM and LR models. Both methods achieve their greatest detection power when risk factors are uncorrelated.

### 3.2.7 Impact of missing risk factors

Figure 14a compares the power for AIM under the three scenarios of no missing factors, a few strong missing factors, and many weak missing factors. A similar comparison for LR is shown in Figure 14b. We simulated three strong and one hundred weak missing factors, but with the overall effect designed to be the same.

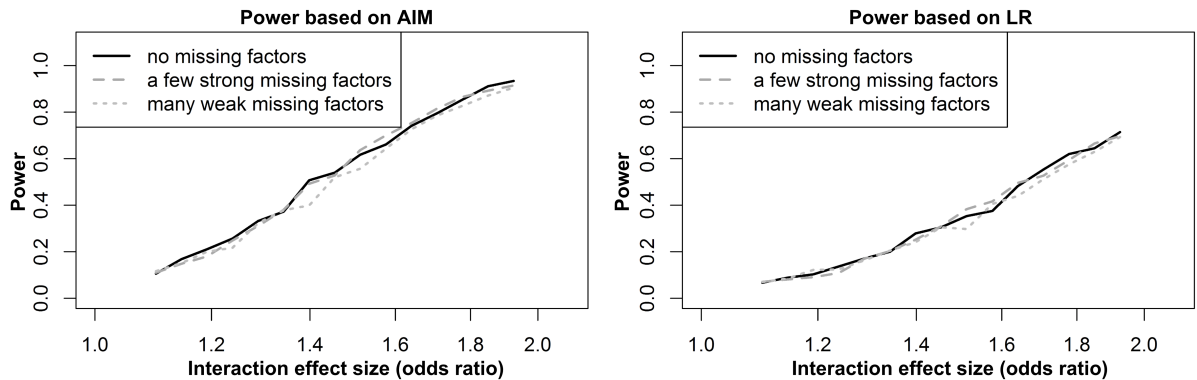

(a) AIM under missing factors  (b) LR under missing factors

Figure 14: Power estimated based on 1000 simulation data sets when there are no missing factors, a few strong missing factors, or many weak missing factors for a) AIM and b) LR.

From these figures we can see that the power does not change too much for both AIM and LR when there are missing factors. The change is so small that we cannot draw definitive conclusions from the two figures; however, the existence of a few strong missing factors does appear to decrease the power. To further assess this, we simulated 10,000 datasets, focusing on an interaction effect size of 1.5, with the results shown in Table 3.

|           | No missing factors   | A few strong missing factors | Many weak missing factors |
|-----------|----------------------|------------------------------|---------------------------|
| AIM power | 0.602 [0.692,0.611]  | 0.555 [0.545, 0.565]         | 0.604 [0.594, 0.614]      |
| LR power  | 0.348 [0.338, 0.357] | 0.317 [0.308,0.326]          | 0.347 [0.337, 0.356]      |

Table 3. Power comparison when there are missing factors. The interaction effect size was fixed at 1.5. Power was estimated based on 10000 simulations. Confidence intervals (shown in brackets) were computed using a binomial distribution.

The existence of a few strong missing factors indeed decreases power for both AIM and LR; however, many weak missing factors did not have any observable effect on the power. Since missing risk factors do not change the power much, it is expected that AIM will still be more powerful than LR, which is confirmed by Figures 15a,b.

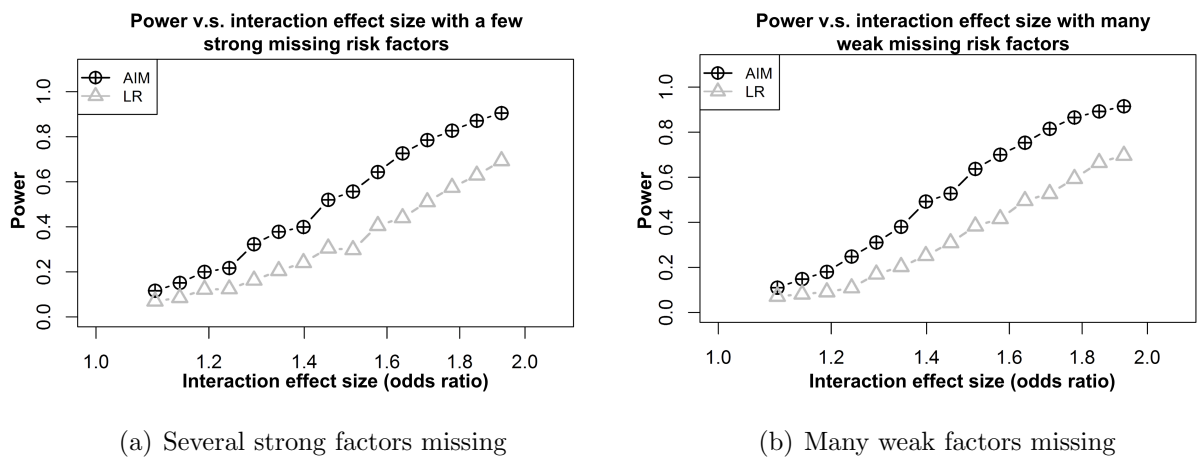

(a) Several strong factors missing    (b) Many weak factors missing

Figure 15: Interaction detection power versus interaction odds ratio for AIM and LR models when a) several strong non-interacting factors are missing and b) many weak non-interacting factors are missing.

### 3.2.8 Impact of surrogate factors

Figure 16 shows how power is affected when surrogate rather than causal risk factors are measured. Surrogate factors have very large impact on the power for both AIM and LR. For example, the power for AIM drops from 0.877 to 0.486 when surrogate factors with correlation coefficient of 0.8 to their causal counterparts are observed. Similarly, the power for LR drops from 0.607 to 0.278. No matter how great the power decrease, we see that AIM's power is always greater than LR's. It is also noteworthy that both AIM and LR are symmetric with respect to the correlation coefficient between surrogate and causal factors.

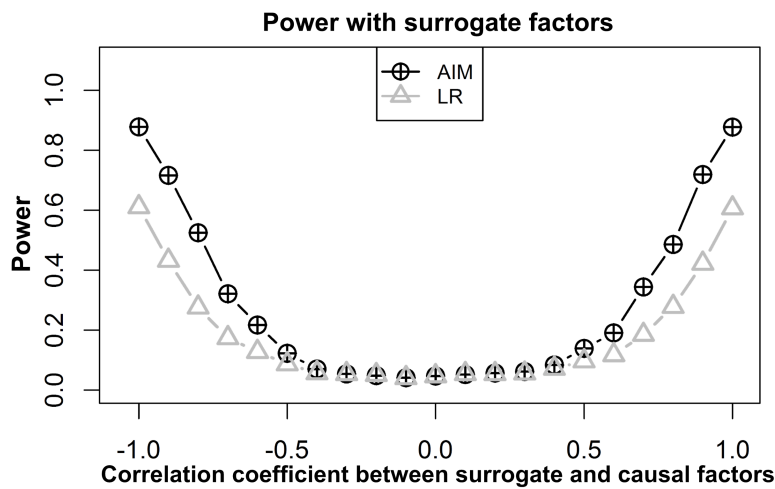

Figure 16: Interaction detection power under the surrogate factors scenario, as a function of the correlation between surrogate and causal factors, for AIM and LR models.

### 3.2.9 Impact of subtypes

Figures 17a,b illustrate how the existence of subtypes impacts the power for both AIM and LR. When there are subtypes, the power for both AIM and LR are significantly increased. It seems that the existence of subtypes make the interaction effect stronger. Even so, we see in Figure 18 that, even with subtypes, AIM has better power to detect the interaction effect than LR.

### 3.2.10 Power for an antagonistic interaction

Figure 19 shows how the power varies with interaction effect size for an interaction that is antagonistic. For a logistic regression model with interaction terms, an antagonistic interaction is defined as one with an interaction coefficient $\beta_3$ that is negative. Equivalently, it is such that the effect size (odds ratio) is smaller than one. The experiments were conducted exactly as in Figure 8 except that the effect size here was set to the reciprocal of the value used to produce Figure 8. The experimental results are consistent with the statement in our theorem, with the power for logistic regression larger than that for AIM when the interaction is antagonistic. However, the reduction in power for AIM relative to LR is quite small. Interestingly, we observe another asymmetric characteristic of AIM through this experiment – logistic regression is symmetric with respect to the log-transformation of the odds ratio, while AIM is not. Indeed, the gain of AIM over LR for synergistic interactions is much larger than the loss relative to LR for antagonistic ones.

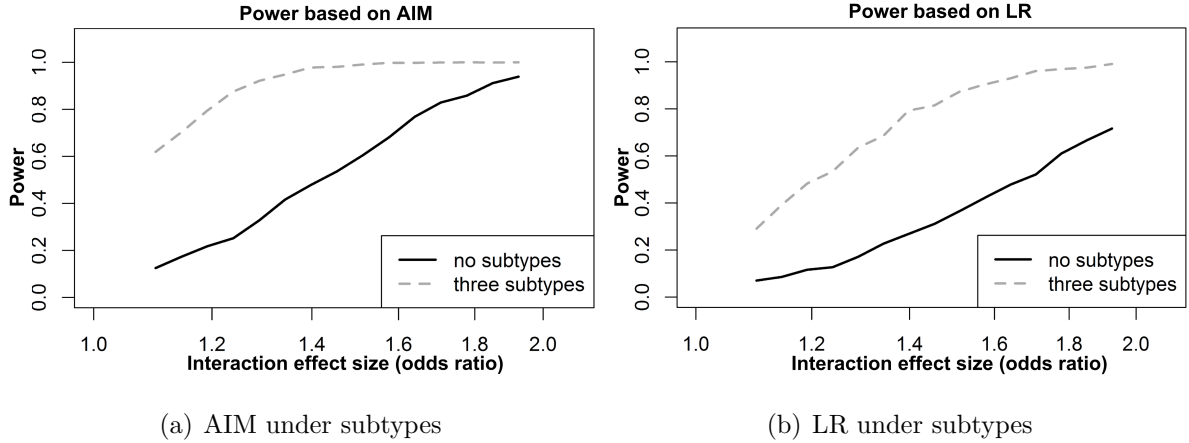

(a) AIM under subtypes  (b) LR under subtypes

Figure 17: Interaction detection power versus interaction odds ratio for a) AIM and b) LR models under the cases of no disease subtypes and three disease subtypes.

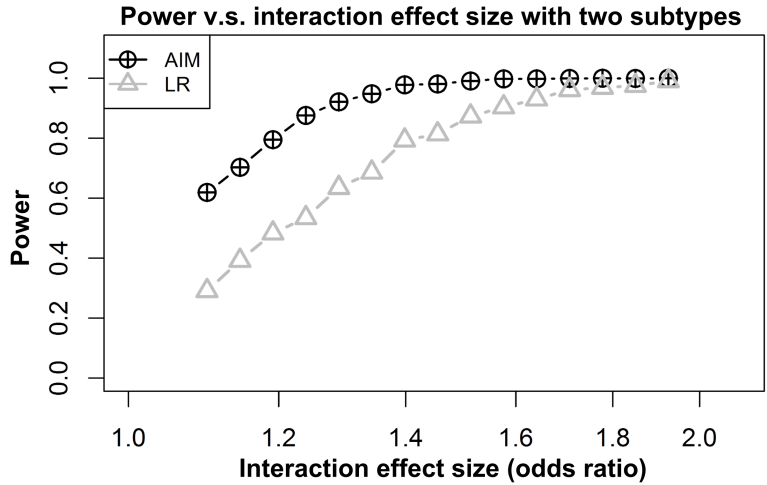

Figure 18: Interaction detection power when there are two disease subtypes, as a function of interaction odds ratio, for AIM and LR models.

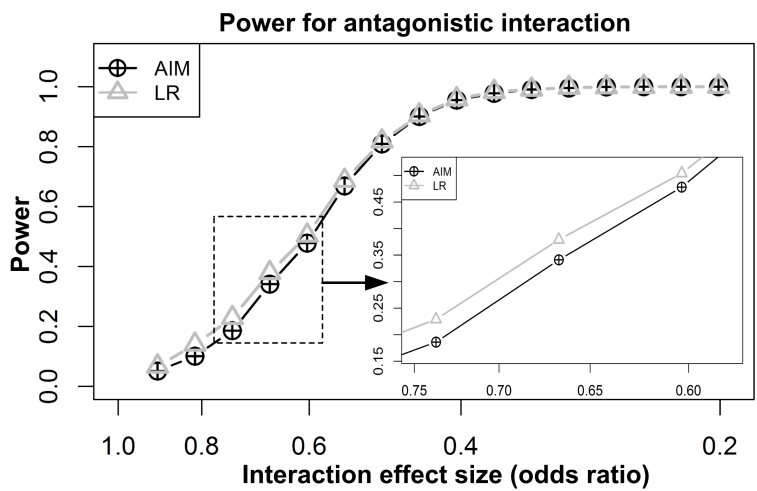

Figure 19: Power of AIM and LR versus interaction effect size for an antagonistic interaction.

### 3.2.11 Experiments on five previous published interaction models

We have also evaluated power to detect interactions for a simulated genomics data set (Li(2011)), derived from real single nucleotide polymorphism (SNP) study data, as part of the New York City Cancer Control Project. The simulated population retained the basic patterns of linkage disequilibrium, missing data, and allele frequencies observed in the original genome scan data. Multiple interactions (five) simultaneously exist in the simulated population (reasonable, considering complex disease mechanisms) and jointly decide the phenotype for each individual. The five interaction models vary in interaction order (from two-way up to five-way), genetic models (dominant, recessive, or additive), incomplete/complete penetrance, minor allele frequency, and marginal effects size. The chosen models, specified in (Li(2011)), were motivated by complex genetic traits (such as autoimmune diseases, diabetes, and arthritis) where there are multiple loci contributing to disease risk and where there are both some relatively large interaction effects as well as more modest ones (Li(2011)). We considered data sets with 1000 samples and 100 SNP variables. Fifteen of these SNPs participate in the five interaction models, with the remaining SNPs having no ground-truth association with the disease status. Figure 20 shows the statistical significance of each ground-truth interaction model, as detected by the AIM and LR models. Note that AIM achieves smaller p-values for all five models. Smaller p-values imply fewer data subjects are needed to detect an interaction at a minimum required level of significance.

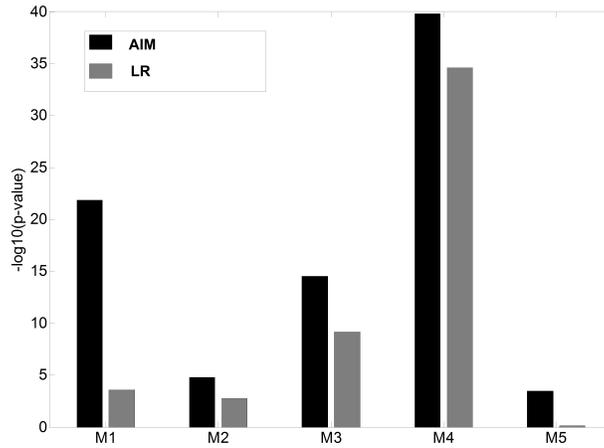

Figure 20: Statistical significance (log p-values) of five ground-truth interactions, as detected by the AIM and LR models.

## 3.3 Overall conclusions from power comparisons

Summarizing the previous discussions, we have the following overall conclusions on the power comparisons: (1) For synergistic interactions, under all scenarios and parameter combinations, AIM is always better powered than LR. (2) The power gain of AIM over LR is larger when the interaction effect size is small and/or when the sample size is relatively small. (3) The power gain of AIM over LR is larger when the main effect is large. (4) The power gain of AIM over LR is larger for small case fractions, as opposed to large case fractions. (5) The existence of missing risk factors, surrogate factors or subtypes may decrease or increase the power for both AIM and LR, but AIM always has larger power than LR. (6) AIM is not only conceptually asymmetric (with respect to disease status) – its power is also asymmetric, with respect to both the case-control ratio and the allele frequency.

# 4 Experiments on Real Datasets

**Interaction between ALDH2 gene and alcohol consumption on esophageal cancer**

Both the ALDH2 gene and alcohol consumption are known factors associated with esophageal cancer. Heavy alcohol consumption has been found to be a risk factor for esophageal cancer in many epidemiological studies (Allen(2009)). When alcohol is metabolized in the liver, it is broken down to acetaldehyde, which is oxidative and recognized as a carcinogen by binding to cellular protein and DNA. The majority (99%) of the produced acetaldehyde is eliminated by the liver. The ALDH2 protein is responsible for degrading the remaining carcinogen. There is a functional polymorphism in the ALDH2 gene, namely ALDH2 Glu478Lys. The Glu allele encodes a protein with normal catalytic activity, while the Lys allele encodes an inactive protein. A defect in the ALDH2 genes significantly reduces the capacity to degrade acetaldehyde and hence exposes an individual to more acetaldehyde than normal. It is biologically plausible for the ALDH2 protein and alcohol consumption to interactingly influence the risk of esophageal cancer (Lewis(2005),Matsuo(2001)). Figure 21 shows the re-analysis of the interaction between the ALDH2 gene and alcohol consumption.

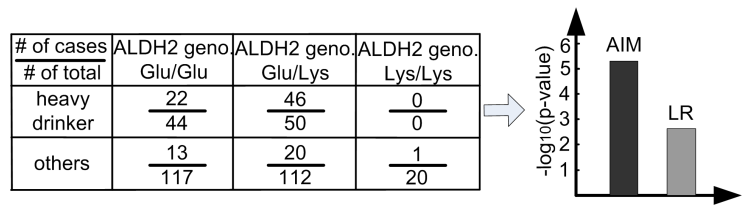

Figure 21: Re-analysis of the interaction between the ALDH2 gene and alcohol consumption.

The data was collected from the first study of the ALDH2-alcohol interaction effect on esophageal cancer. The original report discovered the interaction effect via LR, which was confirmed by follow-up studies (Lewis(2005)) to be a true interaction. The distribution of the cases and the controls are presented in the figure. We re-analyze the data using AIM. The significance through our model is 7.4e-6, compared to a p-value of 2.5e-3 with LR, an almost thousand-fold improvement. There are in total 343 subjects in the study. When all the frequencies of the risk factors and the effect size are kept the same, we estimate that, to achieve the 0.05 significance level, LR requires 142 subjects while AIM needs only 64 subjects.

**Interaction between thrombophilic mutations and oral contraceptive on the venous thrombosis**

The interaction of thrombophilic mutations with oral contraceptives on venous thrombosis is a pronounced example of a gene-environment interrelationship study. A venous thrombosis is a blood clot that forms within a vein, especially in the deep veins of the legs or in the pelvic veins. There are both genetic and environmental risk factors. The R506Q mutation of factor V and the G20210A mutation of prothrombin are two thrombophilic genetic factors (Rosendaal(1999)), moderately common in whites with frequencies of 5% and 2%, respectively. Factor V is a protein of the coagulation system and factor Va (activated factor V) is a highly procoagulant cofactor in the generation of thrombin, which is a crucial element in blood clotting. The R506Q substitution in factor V involves one of three sites that are cleaved by activated protein C. This mutation slows down the proteolytic inactivation of factor Va, which in turn leads to the augmented generation of thrombin (Seligsohn(2001)). Prothrombin is proteolytically cleaved to form thrombin. The G20210A mutation in the 3' untranslated region of the prothrombin gene is associated with an increased level of plasma prothrombin, promoting the generation of thrombin and impairing the inactivation of factor Va by activated protein C (Seligsohn(2001)). The use of oral contraceptives has long been recognized as a risk factor for venous thrombosis. Oral contraceptive has significant effect on the generation of thrombin, by both decreasing the level of factor V and increasing the level of prothrombin.

The interaction between thrombophilic mutations and oral contraceptive is well-established, with multiple epidemiological and mechanical studies (Legnani(2002), Martinelli(1999), Rosing(1999), Vandenbroucke(1994)). Table 4 and Table 5 show two studies illustrating the interaction between the thrombophilic genetic mutation and the use of oral contraceptive.

| thrombophilic genetic risk mutation | oral contraceptive | controls | cases | odds ratio |
|---|---|---|---|---|
| - | - | 444 | 118 | 1 |
| - | + | 166 | 86 | 1.95 |
| + | - | 33 | 42 | 4.79 |
| + | + | 7 | 51 | 27.4 |

Table 4. Legnani et al. study: risk of venous thrombosis according to the presence of thrombophilic genetic mutation and the use of oral contraceptive.

In the Legnani et al. study, the odds ratio associated with the use of oral contraceptive but no thrombophilic genetic risk mutation is 1.95, and the odds ratio associated with genetic defects but no use of contraceptive is 4.79. According to the multiplicative model, the odds ratio associated with the presence of both risk factors should be 9.34, while the observed odds ratio is 27.4. This is strong evidence of interaction. Indeed, by applying LR, we get a p-value of 0.021, which is statistically significant. If we apply AIM, we get a p-value of 0.00062. There are 947 subjects in the Legnani et al. study. When all the frequencies of the risk factors and the effect size are kept the same, we estimate that, to achieve the 0.05 significance level, LR requires 676 subjects, while AIM needs only 303 subjects.

| thrombophilic genetic risk mutation | oral contraceptive | controls | cases | odds ratio |
|---|---|---|---|---|
| - | - | 127 | 35 | 1 |
| - | + | 41 | 52 | 4.60 |
| + | - | 7 | 5 | 2.59 |
| + | + | 4 | 20 | 18.1 |

Table 5. Martinelli et al. study: risk of venous thrombosis according to the presence of thrombophilic genetic mutation and the use of oral contraceptive.

For the Martinelli et al. study, the odds ratio associated with the presence of both risk factors (according to a multiplicative model) is expected to be 11.9, compared to the observed value of 18.1. Both studies have the same effect direction, that is, the observed odds ratio is larger than the expectation. Due to the limited sample size, the conclusion is not statistically significant in the Martinelli et al. study. The p-value generated by LR is 0.618 and the p-value obtained from AIM is 0.183. To achieve the 0.05 significance level, the estimated sample size associated with LR is 4391, while AIM requires just 614 subjects.

**Interaction between NAT2 gene and smoking on bladder cancer**

With hundreds of thousands of new cases diagnosed each year worldwide, bladder cancer is increasingly important for public health, with tobacco smoking the predominant known risk factor. In Europe, smoking is estimated to cause over half of bladder cancer cases in men and one-third of cases among women (Zeegers(2000)). Multiple carcinogens have been found in tobacco smoke, including polycyclic aromatic hydrocarbons, N-nitrosamines, aromatic amines, heterocyclic amines, and aldehydes. Originally inert, these carcinogens may undergo both activation and detoxification. Imbalance between activation and detoxification will increase the bladder cancer risk through accumulation of active carcinogen metabolites and increased DNA adduct formation (Gu(2005)).

The NAT2 gene encodes an enzyme that functions to both activate and deactivate arylamine and hydrazine drugs and carcinogens (Sanderson(2007)). The NAT2 enzyme is particularly active in the liver, gastrointestinal tract, and urinary bladder, among other organs and tissues. Due to the metabolic rate of exogenous compounds, the polymorphisms in the NAT2 gene can be classified into two types – rapid acetylator and slow acetylator. NAT2 slow acetylator is very common in the Caucasian population, estimated to be around 55%. The association of the NAT2 slow acetylator with bladder risk is quite well established, serving as an outstanding example prior to the GWAS era for the replicated association between common genetic polymorphisms and complex diseases.

Multiple studies have consistently shown the interaction between the NAT2 gene and smoking on bladder cancer (Garcia-Closas(2005),Gu(2005),Sanderson(2007)). Table 6 presents the non-meta-analysis study with the largest sample size (Garcia-Closas(2005)). Choosing the bladder cancer risk for "never smoked" and NAT2 fast acetylator as the reference, the odds ratio associated with "smoked before" (i.e., an individual who has smoked before) and NAT2 fast acetylator is 1.86, and the odds ratio associated with "never smoked" and NAT2 slow acetylator is 0.91. According to the multiplicative model, the odds ratio associated with the presence of both risk factors should be 1.69, while the observed odds ratio is 2.89. So the interaction is evident. Indeed, by applying LR, we get a p-value of 0.015. When we apply AIM, we get a p-value of 0.0011. There are 2264 subjects in the study. When all the frequencies of the risk factors and the effect size are kept the same, we estimate that, to achieve the 0.05 significance level, LR requires 1449 subjects and AIM needs 796 subjects.

| NAT2 acetylation genotype | smoking status | controls | cases | odds ratio |
|---|---|---|---|---|
| Fast | never | 131 | 66 | 1 |
| Fast | ever | 362 | 340 | 1.86 |
| Slow | never | 199 | 91 | 0.91 |
| Slow | ever | 438 | 637 | 2.89 |

Table 6. Joint association of tobacco smoking status and NAT2 acetylation genotype with bladder cancer risk.

**Interaction between tobacco smoking and alcohol drinking on esophageal cancer**

It has long been suggested that tobacco smoking and alcohol consumption interplay to influence the risk of cancer (Garro(1990)). Alcohol may act as a cocarcinogen and enhance the carcinogenic effects of other chemicals from tobacco smoking. Indeed, quite a few epidemiological studies have confirmed their interaction effect on esophageal cancer (Castellsague(1999),Lee(2005)). The Castellsague et al. report (Castellsague(1999)) is probably the first large scale case-control study implying the interaction effect of tobacco smoking and alcohol consumption on esophageal cancer. The study showed that the combination of the two factors significantly increased disease risk more than either of them separately. However, although the report demonstrated statistical evidence for both the female group and the all subjects group, it failed to find a significant interaction in the male group. By applying the new model (AIM), interaction analysis generates consistent results. Table 7 presents the subject distribution specified by the status of alcohol drinking, tobacco smoking and esophageal cancer in the Castellsague et al. study. The data are divided into three groups – males, females, and all subjects. In each group, we calculate the interaction effect based on LR and AIM. We can see that the new model consistently generates smaller p-values than LR. In the males group, the p-value is 5.43e-6 based on the new model, while it is 0.81 for LR and far from being considered as significant. We also estimate the sample sizes required for the two models to achieve the 0.05 significance level, again assuming that all the frequencies of the risk factors and the effect size are kept the same. In the males group, LR needs 131413 subjects, compared to 374 subjects required for AIM. In the females group, LR needs 339 subjects and AIM needs 235. In the all group, 596 subjects are necessary for LR , while 312 subjects are sufficient for AIM.

**Summary of results on real data sets:**

1. On esophageal cancer, we re-analyzed the established interaction between ALDH2 gene and alcohol consumption with a more convincing p-value 7.4e-6, compared to 0.0025 for logistic regression. The sample size required for good power is reduced from 142 to 64. [Reanalysis : more convincing statistical evidence, gene-environment interaction]

2. On venous thrombosis, we re-analyzed the established interaction between thrombophilic mutations and oral contraceptive with a more convincing p-value 0.00062, compared to 0.021 for logistic regression. The sample size required for good power is reduced from 676 to 303. [Reanalysis : more convincing statistical evidence, gene-environment interaction]

3. On bladder cancer, we re-analyze the established interaction between NAT2 gene and smok-

| alcohol | smoking | males | | | females | | | all | | |
|---|---|---|---|---|---|---|---|---|---|---|
| | | controls | cases | odds ratio | controls | cases | odds ratio | controls | cases | odds ratio |
| never | never | 189 | 8 | 1 | 234 | 83 | 1 | 423 | 91 | 1 |
| never | ever | 298 | 61 | 4.84 | 55 | 27 | 1.38 | 353 | 88 | 1.16 |
| ever | never | 144 | 24 | 3.94 | 63 | 29 | 1.30 | 207 | 53 | 1.19 |
| ever | ever | 777 | 562 | 17.1 | 19 | 36 | 5.34 | 796 | 598 | 3.49 |
| Logistic regression ($p$) | | 0.81 | | | 0.014 | | | 5.10e-5 | | |
| Log regression ($p$) | | 5.43e-6 | | | 0.0031 | | | 2.11e-8 | | |

Table 7. Joint association of alcohol drinking and tobacco smoking statuses with esophageal cancer risk.

ing, with a more convincing p-value 0.0011, compared to 0.015 through logistic regression. The sample size required for good power is reduced from 1,449 to 796. [Reanalysis : more convincing statistical evidence, gene-environment interaction]

4. On environment-environment interaction between tobacco smoking and alcohol drinking on esophageal cancer, we re-analyzed the data, eliminating originally conflicting/inconsistent results. [Reanalysis : eliminating inconsistency, environment-environment interaction]

5. Across all of our real data set experiments, AIM demonstrated enhanced power compared to LR. We further checked the types of interactions and found that they are all synergistic – in all of these applications, carrying double risk factors engendered larger risk than expected from any individual risk factor. These experimental results thus corroborate our Theorem 4 on the comparison of power.

# 5 Related Work and Discussion

Identification of statistical interactions between participating factors has many practical implications. For instance, significant efforts have been made to investigate gene-environment interactions, as it is well accepted that multiple genetic and environmental factors acting in interconnected biological pathways or networks contribute to the susceptibility and progression of complex human diseases. Besides revealing the mechanisms underpinning the disease, the identification of gene-environment interactions may assist the design of targeted therapies, interventions, or preventive strategies for complex diseases. After all, the genetic variants that are most easily translated for public health or clinical utility will be those that have an obvious corresponding environmental modification identified to be capable of altering the disease risk. In contrast to the impressive accumulation of data resources due to the successful launch of GWAS during the past five years, the progress on analytical approaches has been limited, with LR still the de facto standard (Cordell(2009)). Despite its popularity, LR occasionally receives critiques, albeit mainly for its lack of power or due to its excessively large required sample sizes. However, the fundamental problems we have discussed in this paper have received little attention. Various efforts have been made to enhance LR's power. One major school is based on the case-only study (Cordell(2009)). Under two assumptions – (1) the two risk factors are independent in the population and (2) the disease incidence rate is rare – it can be shown that LR for interaction analysis reduces to association analysis between the two risk factors in the case group only. Simulation studies have demonstrated larger power of case-only studies compared to LR. However, it was pointed out that the power gain was purely owing to these strong assumptions, since, under them, fewer parameters need to be estimated and the true interaction effect can be more easily distinguished from the null distribution with fewer degrees of freedom. Yet, in real applications, these assumptions may not hold. A gene may influence the environment to which an individual is exposed. For example, genetic makeup may be a strong determinant of lifestyle. Even for some seemingly unlikely dependency, the independence assumption can be violated indirectly, for instance, through family history. For example, a potential inheritor of the BRCA1 gene may tend to opt for oral contraceptives because of the history of breast cancer in the family. Violation of the independence assumption often leads to inflated false positive rates. Sometimes it will also result in decreased power. (Wu(2011)) provides such an empirical example. The case-only method missed the interaction between the ALDH2 gene and drinking status on esophageal squamous-cell carcinoma, whereas standard LR successfully detected it, because a person with the risk allele in the ALDH2 gene tends to not drink due to the flushing reaction while drinking.

New variants of the case-only method, including empirical Bayes (Mukherjee(2008)) and model averaging (Li(2009)), were proposed to combine the strengths of the case-only study and of LR, aiming at gaining power by exploiting the assumption of independence and yet protecting against false positives when the independence assumption is violated. However, these methods are essen-tially weighted averages of the case-only and case-control statistics, and hence they are necessarily liberal under the violation of the independence assumption. At the same time, case-only and its variants will be invalid for common diseases like diabetes or heart diseases due to the violation of the assumption of rare incidence. More importantly, both the case-only method and its variants share the same principle with LR, that is, that null models be multiplicative for disease risks coming from multiple factors. Therefore, the fundamental problems we discussed pertaining to LR also apply to the case-only method and its variants.

All the approaches discussed above – LR, the case-only method, and its variants – can be considered multiplicative models because their null hypotheses all posit a product of disease risks when all risk factors are present. An alternative is the so-called additive model, which hypothesizes an additive effect of disease risks when multiple risk factors are present. This model was highly motivated by the public health goal of finding cost-effective intervention strategies for disease reduction, since departures from additive risk would identify special groups that benefit most from a given intervention. On the one hand, an interaction relevant to the public health goal is not necessarily the most biologically meaningful. On the other hand, we do realize that the additive model can be derived as a special case of AIM if we assume the phenotype we are interested in, that is, the case, has low prevalence in the population so that mathematically $\log(1 - p) \approx -p$. Nonetheless, there are disease domains for which this approximation is wildly violated. One is common diseases. For instance, according to the Centers for Disease Control and Prevention (CDC), the prevalence of coronary heart disease is as high as 19.8% among persons $\geq 65$ years in 2010. According to the American Diabetes Association, 11.3% of all people $\geq 20$ years have diabetes in 2011. If the population is restricted to those $\geq 65$ years, the prevalence reaches 26.9%. It is also highly likely that a rare disease becomes quite common when referring to a special population, such as Finnish heritage disease. Another important scenario is where the interest of a study is on the progression of the disease instead of its occurrence. Even though the incidence rate can be very low, the poor prognosis group, which is often considered as the 'cases', can constitute *any* fraction of the whole patient group.

# 6 Conclusions

We first considered the widely used logistic regression model, identifying its limitations as a plausible model for disease risk. We further identified that logistic regression is not theoretically supported for hypothesis testing on statistical interactions between risk factors under the following common scenarios: 1) when there are additional (unmeasured) risk factors; 2) when measured factors are "surrogates", imperfectly correlated with the true factors; 3) when there are multiple disease subtypes. Alternatively, we proposed as the null the Asymmetric Independence Model (AIM) which: i) crucially, unlike LR, is *asymmetric* with respect to "diseased" and "healthy" statuses; ii) more generally, does comport with well-accepted biological models; and iii) whose mathematical form is preserved under all of the above confounding scenarios. Finally, we gave a precise, operational definition of a "synergistic" interaction, an interaction type commonly encountered in practice, for which we proved mathematically that AIM has greater detection power than LR, *irrespective of* whether the above confounding scenarios are active. Experiments evaluating AIM and LR both on simulated data sets as well as on four real disease case-control study domains demonstrate AIM's improved detection power over LR. Moreover, controlled experiments demonstrate both the inflated type 1 error of LR, and the type 1 error resilience and better detection power of AIM, under unmeasured, surrogate factor, and disease subtype scenarios. Through simulation studies, we also characterized how, for each of the two methods, power depends on an array of population variables and experiment design parameters, including sample size, effect size, case-control ratio, risk factor allele frequency, p-value threshold, main effects, correlation between risk factors, missing factors, surrogate factors, and disease subtypes. Beyond observing that AIM achieved improved power over LR under all the tested scenarios involving synergistic interactions, some of our interesting findings include that: 1) The power gain of AIM over LR is larger when the interaction effect size is small and/or when the sample size is relatively small; 2) The power gain of AIM over LR is larger when the main effect is large; 3) Fitting to its name, AIM is not only conceptually asymmetric – its power is *also* asymmetric, with respect to the case-control ratio, the allele frequency, and the log of the odds ratio; 4) While LR does have a detection power advantage over AIM for antagonistic interactions, we observed very modest power differences between the two models for the simulated antagonistic interactions we investigated, for all tested interaction effect sizes. We also note that, in many instances, it may be possible to determine the hypothesized interactions "direction" (synergistic or antagonistic) and choose the null hypothesis model accordingly.

# 7 Supplementary

**Appendix A: Model Inconsistency of LR in the Presence of Surrogate Risk Factors**

Consider an example where $N = 2$ and where, when both causal factors are observed, the baseline LR model has parameters $\beta_0 = -4$, $\beta_1 = 2$, and $\beta_2 = 2$. The corresponding disease distribution is shown in Table 8.

| p | X2 = 0 | X2 = 1 |
|---|---|---|
| X1 = 0 | 0.0180 | 0.1192 |
| X1 = 1 | 0.1192 | 0.5000 |

Table 8. Posterior probability of disease with two causal SNPs $X_1$ and $X_2$ under the logistic regression model.

However, suppose now that, rather than observing the causal factors $X_1$ and $X_2$, we observe two surrogate factors, $X_1'$ and $X_2'$, correlated with their respective causal factors and statistically independent of each other. Assume $P(X_m' = 0|X_m = 0) = 0.9$, $P(X_m' = 1|X_m = 1) = 0.9$, and let $P(X_m = 0) = 0.5, m = 1, 2$.

Applying Bayes rule and using the fact, for this example, that $P(X_1 = i, X_2 = j) = P(X_1' = i', X_2' = j') = 0.25, \forall i, j, i', j'$, we find that

$$P_{LR}(C = 1|X_1' = k_1, X_2' = k_2) = \qquad (8)$$
$$\sum_{i=0}^{1}\sum_{j=0}^{1} P_{LR}(C = 1|X_1 = i, X_2 = j)P(X_1' = k_1, X_2' = k_2|X_1 = j, X_2 = j).$$

The resulting disease distribution is shown in Table 9.

Now, assuming that the LR model form is preserved when the surrogate factors, rather than the causal factors, are observed, let us denote by $\beta_0', \beta_1'$, and $\beta_2'$ the parameter values for the new LR model. We can compute these based on Table 9, as follows:

$$\beta_0' = \log\left(\frac{P(C = 1|X_1' = 0, X_2' = 0)}{1 - P(C = 1|X_1' = 0, X_2' = 0)}\right) = \log\left(\frac{0.0410}{1 - 0.0410}\right) = -3.1523$$

$$\beta_1' = \log\left(\frac{P(C = 1|X_1' = 0, X_2' = 1)}{1 - P(C = 1|X_1' = 0, X_2' = 1)}\right) - \beta_0' = \log\left(\frac{0.1444}{1 - 0.1444}\right) + 3.1523 = 1.3731$$

|   p    | X2' = 0 | X2' = 1 |
|--------|---------|---------|
| X1' = 0 | 0.0410  | 0.1444  |
| X1' = 1 | 0.1444  | 0.4266  |

Table 9. Posterior probability of disease with two surrogate variables $X_1'$ and $X_2'$ under the logistic regression model.

$$\beta_2' = \log\left(\frac{P(C=1|X_1'=1, X_2'=0)}{1-P(C=1|X_1'=1, X_2'=0)}\right) - \beta_0' = \log\left(\frac{0.1444}{1-0.1444}\right) + 3.1523 = 1.3731$$

The odds for the genotype $(X_1' = 1, X_2' = 1)$ based on this model is:

$$\frac{P(C=1|X_1'=1, X_2'=1)}{1-P(C=1|X_1'=1, X_2'=1)} = e^{\beta_0' + 2\beta_1'} = 0.6662. \tag{9}$$

However, from Table 9, the odds for $(X_1' = 1, X_2' = 1)$ is $\frac{0.4266}{1-0.4266} = 0.7440$. Thus, the assumption that the LR form is preserved when surrogate factors, rather than the true causal factors are observed (that is, that the correlation structure between the causal and surrogate factors can be accounted for while preserving the baseline LR model form) is contradicted in this example and, thus, as a general rule.

**Appendix B**

**Theorem 5: AIM's log-likelihood function is concave in its model parameters.**

*Proof:* The Hessian matrix of second order partial derivatives $H = [\partial^2 \log L / \partial \beta_i \partial \beta_j] = -\sum_{i=1}^{M} I_i w_i \underline{y}_i \underline{y}_i^T$, where $\underline{y}_i = [1 \; \underline{x}_i]$, $I_i$ is an indicator with value 1 if subject $i$ is a case and zero otherwise, and $w_i = P_{\text{AIM}}(C=1|\underline{x}_i)/(1-P_{\text{AIM}}(C=1|\underline{x}_i))^2 > 0$. A non-negatively weighted outer product is non-negative definite, and a sum of non-negative definite matrices is also non-negative. Thus, the Hessian matrix (with a negative sign out front) is non-positive definite. Further, assuming the matrix is full rank, it is negative definite. Thus, the log-likelihood function is concave in its parameters.

## Appendix C: Constrained Maximum Likelihood Algorithm for Estimating the AIM Model

As shown in Appendix B, AIM's log-likelihood objective function is concave in the parameters $\underline{\beta}$. Since the linear constraints on these parameters form a convex feasible region in the parameter space, constrained MLE of AIM's model parameters amounts to a convex optimization (Boyd(2003)), for which globally optimal parameter estimates (or those approximating the global optimum to any required level of precision) can be efficiently found. In the sequel, we describe the two-step algorithm (Boyd(2003)), based on use of Newton-Raphson as initialization and a log-barrier interior point algorithm (Boyd(2003)) to refine the solution, which we used in producing AIM MLE parameter estimates.

Assume $M$ subjects, with $I_i = 1$ indicating the $i$-th subject is a case and $I_i = 0$ indicating a control. Define the augmented factor vector $\underline{y} = [1 \ \underline{x}^T]$. Thus, given the parameter vector $\underline{\beta}$, the class posterior log-likelihood over all M subjects (assuming subject independence) is:

$$L(\underline{\beta}) = \sum_{i=1}^{M} \left( (1 - I_i)\underline{\beta}^T \underline{y}_i + I_i \log(1 - e^{\underline{\beta}^T \underline{y}_i}) \right). \tag{10}$$

Maximum likelihood estimation for $\underline{\beta}$ is posed as:

$$\arg\max_{\underline{\beta}} L(\underline{\beta}) \text{ subject to } \underline{\beta}^T \underline{y}_i \leq 0, i = 1, \ldots, M. \tag{11}$$

Note that even if $M > 2^N$, the number of distinct constraints is upper-bounded by the number of unique binary factor vectors, $2^N$.

**Optimization Strategy:**

An important empirical observation is that, frequently, the solution to the unconstrained problem (obtained by ignoring the constraints in (11)) in fact satisfies all the constraints. We thus propose a two-stage optimization exploiting this to achieve computational efficiency in practice. In the first stage, Newton-Raphson is used to solve the unconstrained MLE problem, albeit with a check on constraint satisfaction after each iteration. If any constraint is violated, we terminate this stage and go to the second stage; otherwise, Newton-Raphson iterations are applied until the specified convergence target is met. In the provisional second stage, we apply the log-barrier method (Boyd(2003)) to solve the constrained problem. The basic idea is to define a penalized log-likelihood function that leaves the log-likelihood unmodified when there are no constraint violations but severely penalizes any violations. The penalty function thus acts as a "barrier", ensuring the parameter vector always remains feasible while iteratively maximizing the penalized objective starting from the interior of the feasible region. Since the penalty function described above is not in general differentiable, we instead construct a differentiable surrogate penalty, indexed by a parameter $t$, that approaches the desired penalty in the limit as $t \to \infty$. The modified penalized log-likelihood (based on the surrogate penalty) is maximized for an increasing sequence of parameter values $t^{(0)}, t^{(1)}, t^{(2)}, \ldots$ via a continuation method, i.e. with the solution for the objective indexed by $t^{(i)}$ used as initialization for $t^{(i+1)}$. From the duality theory of convex optimization (Boyd(2003)), one can strictly bound the log-likelihood deficiency of the current solution (with respect to the global maximum log-likelihood), which inversely depends on $t$. Thus, one can achieve any desired precision to the global maximum by optimizing for $t$ sufficiently large. This log-barrier approach is described in detail in (Boyd(2003)). Below, we summarize the main algorithm steps.

**First stage: Newton-Raphson to solve the unconstrained problem**

Given the current estimate $\underline{\beta}^{(k)}$, the next estimate $\underline{\beta}^{(k+1)}$ is produced by: $\underline{\beta}^{(k+1)} = \underline{\beta}^{(k)} - H(\underline{\beta}^{(k)}) \nabla_{\underline{\beta}} L(\underline{\beta}^{(k)})$, where $H(\cdot)$ is the Hessian matrix:

$$H(\underline{\beta}) = \sum_{i=1}^{M} I_i P_{\text{AIM}}(C=1|\underline{x}_i; \underline{\beta})/(1 - P_{\text{AIM}}(C=1|\underline{x}_i; \underline{\beta}))^2 \underline{y}_i \underline{y}_i^T \qquad (12)$$

and

$$\nabla_{\underline{\beta}} L(\underline{\beta}) = \sum_{i=1}^{M} (1-I_i)\underline{y}_i + I_i P_{\text{AIM}}(C=1|\underline{x}_i; \underline{\beta})/(1 - P_{\text{AIM}}(C=1|\underline{x}_i; \underline{\beta})) \underline{y}_i \underline{y}_i^T. \qquad (13)$$

The Newton method has a quadratic convergence rate. It usually converges in less than 10 iterations in our application.

**Second stage: Barrier method to solve the constrained problem**

Letting $I_+(x) = 0$ if $x > 0$ and $\infty$ otherwise, the preceding constrained optimization problem is equivalent to: $\max_{\underline{\beta}} L(\underline{\beta}) + \sum_{i=1}^{M} I_+(-\underline{\beta}^T \underline{y}_i)$. Note that the optimum must occur for $\underline{\beta}$ satisfying $\underline{\beta}^T \underline{y}_i < 0, i = 1, \ldots, M$ to avoid the infinite penalty. Moreover, in this (feasible) case, the penalized log-likelihood reduces to the pure log-likelihood. Since $I^+(x)$ is not differentiable, we substitute the following penalty function that approaches $I^+(x)$ in the limit of large t: $\phi_t(x) = \frac{1}{t} \log(x)$ for $x > 0$ and $\infty$ otherwise. For given $t$, we thus solve: $\max_{\underline{\beta}} L(\underline{\beta}) + \sum_{i=1}^{M} \phi_t(-\underline{\beta}^T \underline{y}_i)$. Note that this objective function is also concave. Thus, for each $t$, we can apply the Newton method for its maximization. As $t \to \infty$, this modified problem approaches the original problem.

**Schedule for $t$ and choice of $t^{(0)}$:**

The control parameter is updated using an exponential schedule: $t^{(k+1)} = \lambda t^{(k)}$, $\lambda > 1$. As discussed in (Boyd(2003)), a reasonable choice for $t^{(0)}$ is such that $M/t^{(0)}$ is approximately $\lambda(L^* -$ $L(\underline{\beta}^{(0)})$, where $L^*$ is the true maximum. Since $L^*$ and $L(\underline{\beta}^{(0)})$ are both unknown, we approximate their difference by the log-likelihood difference between the last two Newton iterations in stage 1 (immediately prior to detecting constraint infeasibility).

While large $\lambda$ reduces the number of optimizations performed in reaching a target value $t_{\text{final}}$, a "too large" difference $(t^{(k+1)} - t^{(k)})$ may mean that the solution at step $k$ gives poor initialization at step $k+1$, translating to slow convergence of the Newton algorithm. There is thus a tradeoff in the choice of $\lambda$. Experiments suggest that values in the range $[3, 100]$ are reasonable choices. In our experiments we set $\lambda = 20$.

**Parameter Choices:**

For the inner loop (Newton minimization), we stop if the increase in log-likelihood is less than $10^{-6}$. For the outer loop (over $t$), we stop if $t^{(k)} > 10^6 M$. From the dual optimization theory (Boyd(2003)), this ensures a log-likelihood deficiency of less than $10^{-6}$.

**Appendix D: Proof of AIM's Consistency Under Missing Factors**

Suppose that the $N$ factors $X_i, i = 1, \ldots, N$, each with discrete range denoted $\mathcal{R}(X_i)$, are statistically independent and that, given all factors observed, the disease status is generated according to the AIM posterior $P_{\text{AIM}}(C = 1|\underline{x}) = 1 - P(C = 0|\underline{x}) = 1 - e^{\beta_0 + \sum_{i=1}^{N} \beta_i x_i}$, where $\beta_0$ will be referred to as the background parameter. We prove that the posterior on disease status remains of this form when only a subset of the factors are observed. Let $\mathcal{S} \subset \{1, 2, \ldots, N\}$ be the indexes of the observed factors. We thus show that, for any $\mathcal{S}$, $\log(1 - P(C = 1|\underline{x})) = e^{\beta'_0 + \sum_{i \in \mathcal{S}} \beta'_i x_i}$, where, *moreover*, $\beta'_i = \beta_i, i \in \mathcal{S}$.

Let $\mathcal{S}_j$ be any subset of cardinality $N - j$. The proof is by induction on $j$. First, note that $\mathcal{S}_0 = \{1, 2, \ldots, N\}$ and, since by assumption the posterior is the AIM model when all factors are observed, for $\mathcal{S}_0$ the posterior is indeed of the AIM form, with $\beta'_i = \beta_i, i = 1, \ldots, N$ and $\beta'_0 = \beta_0$. Thus, the results holds at $j = 0$. Next, assume that the result holds for any subset of size $j$, $\mathcal{S}_j = \{i_1, i_2, \ldots, i_j\}$. That is, the posterior on disease status, given observation of the factors in the subset $\mathcal{S}_j$, is of the AIM form, with unperturbed parameter values $\beta'_i = \beta_i, i \in \mathcal{S}_j$. Let us denote the background parameter value in this posterior by $\tilde{\beta}_0$. We must show consequentially the result also holds for the subsets $\mathcal{S}_{j+1}$. Note that if we remove one factor from any subset $\mathcal{S}_j$, we obtain a subset $\mathcal{S}_{j+1}$. We can express the posterior for this subset by:

$$P[C = 0|x_{i_1}, x_{i_2}, \ldots, x_{i_{j-1}}] \tag{14}$$

$$
\begin{aligned}
&= \sum_{x_{i_j} \in \mathcal{R}(X_{i_j})} P[C = 0, x_{i_j} | x_{i_1}, x_{i_2}, \ldots, x_{i_{j-1}}] \\
&= \sum_{x_{i_j} \in \mathcal{R}(X_{i_j})} P[C = 0 | x_{i_1}, x_{i_2}, \ldots, x_{i_{j-1}}, x_{i_j}] P[X_{i_j} = x_{i_j} | x_{i_1}, x_{i_2}, \ldots, x_{i_{j-1}}] \\
&= \sum_{x_{i_j} \in \mathcal{R}(X_{i_j})} P[C = 0 | x_{i_1}, x_{i_2}, \ldots, x_{i_{j-1}}, x_{i_j}] P[X_{i_j} = x_{i_j}] \\
&= \sum_{x_{i_j} \in \mathcal{R}(X_{i_j})} (1 - e^{\tilde{\beta}_0 + \sum_{l=1}^{j} \beta_{i_l} x_{i_l}}) P[X_{i_j} = x_{i_j}] \\
&= 1 - e^{\tilde{\beta}_0 + \sum_{l=1}^{j-1} \beta_{i_l} x_{i_l}} \left( \sum_{x_{i_j} \in \mathcal{R}(X_{i_j})} e^{\beta_{i_j} x_{i_j}} P[X_{i_j} = x_{i_j}] \right).
\end{aligned}
$$

That is the posterior's form is preserved, with $\beta'_{i_l} = \beta_{i_l}, i_l \in \mathcal{S}_{j+1}$ and where we identify the new background parameter, with no dependence on any of the factors, as $\beta'_0 = \tilde{\beta}_0 + \log(\sum_{x_{i_j} \in \mathcal{R}(X_{i_j})} e^{\beta_{i_j} x_{i_j}} P[X_{i_j} = x_{i_j}])$.

Q.E.D.

**Appendix E: Proof of AIM's Consistency Under Surrogate Factors**

Suppose there are $N$ true disease factors $X_i$, each with discrete range space $\mathcal{R}(X_i), i = 1, \ldots, N$, and with one special value of the range space, denoted $v_i$, corresponding to the disease factor not being *active*[1]. Suppose that, when all $N$ factors are observed, the disease status posterior has the AIM form:

$$\log(1 - P(C = 1 | x_1, x_2, \ldots, x_N)) = \beta_0 + \sum_{i=1}^{N} \beta_i(x_i), \tag{15}$$

where if $x_i = v_i$, $\beta_i(x_i) = 0, \forall i$. Now suppose that there is a subset of factors which are *not* observed. However, rather than being missing, *surrogate* factors, correlated with these true factors, are observed. Let $X'_i$ denote the observed surrogate factor correlated with true factor $X_i$. Assume each true factor $X_i$ is conditionally independent of all other factors (true or surrogate) given its surrogate $X'_i$. Further, assume that the disease status is conditionally independent of a surrogate factor given its true factor. Under these assumptions, we prove that the posterior probability on disease status, given all observed factors (both true and surrogate factors) *remains* of the AIM form and, moreover, is such that, for a true observed factor $X_i = x_i$, its parameter value is $\beta_i(x_i) \forall x_i \in \mathcal{R}(X_i)$, that is the parameter value is *unaltered* by the presence of surrogate factors.

---

[1]Here we are allowing each factor to have a non-binary range space. In the case of binary factors, consistent with the derivation of the AIM model in the main paper, the value indicating a factor's inactivity is $v_i = 0$. More generally, for non-binary factors, we are supposing there is a value $v_i$ indicative of a factor's inactivity.

Let $\mathcal{S}_j = \{i_1, i_2, \ldots, i_j\}$ be any subset of surrogate factors, of cardinality $j$, with companion set $\bar{\mathcal{S}}_j = \{l_1, l_2, \ldots, l_{N-j}\}$. The proof is by induction on $j$. First, note that $S_0 = \{\}$, with all true factors observed. Since by assumption the posterior in this case has the AIM form with parameter values $\beta_i, i = 1, \ldots, N$, the result holds for $j = 0$. Next, assume that the result holds for some $j > 0$, i.e. for the subsets $\mathcal{S}_j$. That is, given $j$ observed surrogate factors and $N - j$ observed true factors, the posterior form is the AIM form, where, further, for each true factor value $X_i = x_i$, its parameter value is $\beta_i(x_i)$, i.e. the same value as when no observed factors are surrogates. Denoting the set of parameter values for a surrogate factor $X'_i$ by $\beta'_i(\omega), \omega \in \mathcal{R}(X'_i)$, the posterior form for any surrogate factor subset $\mathcal{S}_j$ is thus:

$$\log(1 - P(C = 1 | x'_{i_1}, x'_{i_2}, \ldots, x'_{i_j}, x_{l_1}, x_{l_2}, \ldots, x_{l_{N-j}})) = \tag{16}$$
$$\tilde{\beta}_0 + \sum_{m=1}^{j} \beta'_{i_m}(x'_{i_m}) + \sum_{n=1}^{N-j} \beta_{l_n}(x_{l_n}),$$

where $\tilde{\beta}_0$ is the value of the background parameter (which will not in general equal $\beta_0$).

We must show this result consequentially holds for the subsets $\mathcal{S}_{j+1}$. The posterior for a subset $\mathcal{S}_{j+1}$ is:

$$P[C = 0 | x'_{i_1}, x'_{i_2}, \ldots, x'_{i_{j+1}}, x_{l_1}, x_{l_2}, \ldots, x_{l_{N-j-1}}] = \tag{17}$$
$$\sum_{x_{i_{j+1}} \in \mathcal{R}(X_{i_{j+1}})} P[C = 0, x_{i_{j+1}} | x'_{i_1}, x'_{i_2}, \ldots, x'_{i_{j+1}}, x_{l_1}, x_{l_2}, \ldots, x_{l_{N-j-1}}] =$$
$$\sum_{x_{i_{j+1}} \in \mathcal{R}(X_{i_{j+1}})} P[C = 0 | x'_{i_1}, x'_{i_2}, \ldots, x'_{i_{j+1}}, x_{l_1}, x_{l_2}, \ldots, x_{l_{N-j-1}}, x_{i_{j+1}}] \cdot$$
$$P[X_{i_{j+1}} = x_{i_{j+1}} | x'_{i_1}, x'_{i_2}, \ldots, x'_{i_{j+1}}, x_{l_1}, x_{l_2}, \ldots, x_{l_{N-j-1}}] =$$
$$\sum_{x_{i_{j+1}} \in \mathcal{R}(X_{i_{j+1}})} P[C = 0 | x'_{i_1}, x'_{i_2}, \ldots, x'_{i_j}, x_{i_{j+1}}, x_{l_1}, x_{l_2}, \ldots, x_{l_{N-j-1}}] P[X_{i_{j+1}} = x_{i_{j+1}} | x'_{i_{j+1}}] =$$
$$\sum_{x_{i_{j+1}} \in \mathcal{R}(X_{i_{j+1}})} e^{\tilde{\beta}_0 + \sum_{n=1}^{N-j-1} \beta_{l_n}(x_{l_n}) + \sum_{m=1}^{j} \beta'_{i_m}(x'_{i_m})} e^{\beta_{i_{j+1}}(x_{i_{j+1}})} P[X_{i_{j+1}} = x_{i_{j+1}} | x'_{i_{j+1}}] =$$
$$e^{\tilde{\beta}_0 + \sum_{n=1}^{N-j-1} \beta_{l_n}(x_{l_n}) + \sum_{m=1}^{j} \beta'_{i_m}(x'_{i_m})} \left( \sum_{x_{i_{j+1}} \in \mathcal{R}(X_{i_{j+1}})} e^{\beta_{i_{j+1}}(x_{i_{j+1}})} P[X_{i_{j+1}} = x_{i_{j+1}} | x'_{i_{j+1}}] \right) =$$
$$e^{\tilde{\beta}_0 + \sum_{n=1}^{N-j-1} \beta_{l_n}(x_{l_n}) + \sum_{m=1}^{j} \beta'_{i_m}(x'_{i_m}) + \beta'_{i_{j+1}}(x'_{i_{j+1}})}.$$

Here, the third resultant is obtained using the fact that disease status is conditionally independent of $X'_{i_{j+1}}$ given $X_{i_{j+1}}$ and the fact that $X_{i_{j+1}}$ is conditionally independent of all other fac-tors given $X'_{i_{j+1}}$. The fourth resultant is obtained because, by assumption, the result holds for the subsets $S_j$. Finally, in the final result, we have made the identification that $\beta'_{i_{j+1}}(x'_{i_{j+1}}) = \sum_{x_{i_{j+1}} \in \mathcal{R}(X_{i_{j+1}})} e^{\beta_{i_{j+1}}(x_{i_{j+1}})} P[x_{i_{j+1}}|x'_{i_{j+1}}]$, i.e. a quantity that is a function only of $x'_{i_{j+1}}$. Thus, the posterior's form is preserved for subsets of size $j+1$.

Q.E.D.

**Appendix F: Proof of AIM's Consistency Under Disease Subtypes**

Here we will only prove that AIM *is* a consistent model under the scenario of a heterogeneous disease with multiple subtypes. While not shown here, the inconsistency of LR as a model for a heterogeneous disease can be proven by counterexample, just as we have done for the missing and surrogate factor confounding scenarios.

Suppose that there are $K$ disease subtypes, each with a disease subtype status posterior of the AIM form:

$$\log(1 - P(D_k = 1|x_1, x_2, \ldots, x_N)) = \beta_{k0} + \sum_{i=1}^{N} \beta_{ki} x_i, k = 1, \ldots, K. \tag{18}$$

Further, assume that these subtypes are conditionally independent given the observed factors $X_i, i = 1, \ldots, N$. Under these assumptions, we will prove that the heterogeneous disease status $C = \bigcup_{k=1}^{K} D_k$ has a posterior that is also of the AIM form, i.e.

$$\log(1 - P(C = 1|x_1, x_2, \ldots, x_N)) = \beta_0 + \sum_{i=1}^{N} \beta_i x_i. \tag{19}$$

Furthermore, $\beta_i = \sum_{k=1}^{K} \beta_{ki}, i = 1, \ldots, N$ and $\beta_0 = \sum_{k=1}^{K} \beta_{k0}$. That is the strengths of each of the factors for the heterogeneous disease is additive over the strengths for each of the subtypes. We note an important implication of this result: to do inference on the heterogeneous disease using the AIM model, one need not have *any* prior knowledge of how many (and whether in fact) multiple disease subtypes exist for the given disease domain. On the other hand, since the LR form (if assumed to be valid for the subtypes) is not preserved for the heterogeneous disease, such statement will not hold for LR.

The proof of the theorem is as follows. We have $(C = 0) \Leftrightarrow \bigcap_{k=1}^{K}(D_k = 0)$. Therefore, $P(C = 0|x_1, x_2, \ldots, x_N) = P(D_1 = 0, D_2 = 0, \ldots, D_K = 0|x_1, x_2, \ldots, x_N)$. But since the disease subtypes are conditionally independent given the factors, $P(D_1 = 0, D_2 = 0, \ldots, D_K = 0|x_1, x_2, \ldots, x_N) =$

$\prod_{k=1}^{K} P(D_k = 0|x_1, \ldots, x_N)$. Thus,

$$\begin{align}
\log(P(C = 0|x_1, x_2, \ldots, x_N)) &= \sum_{k=1}^{K} \log(P(D_k = 0|x_1, \ldots, x_N)) \tag{20} \\
&= \sum_{k=1}^{K}(\beta_{k0} + \sum_{i=1}^{N} \beta_{ki} x_i) \\
&= (\sum_{k=1}^{K} \beta_{k0}) + \sum_{i=1}^{N} x_i (\sum_{k=1}^{K} \beta_{ki}) \\
&= \beta_0 + \sum_{i=1}^{N} x_i \beta_i,
\end{align}$$

where $\beta_0 = \sum_{k=1}^{K} \beta_{k0}$ and $\beta_i = \sum_{k=1}^{K} \beta_{ki}$.

Q.E.D.

## Appendix G: Proof of AIM and LR Detection Power Relationship for Synergistic Interactions

Suppose there are $N = 2$ binary factors. Let $p_{lm} \equiv P(C = 1|X_1 = l, X_2 = m), l \in \{0,1\}, m \in \{0,1\}$ be the true posterior probability of disease, for each of the four possible factor pair "genotypes". In this Appendix, we will prove that, if $p_{10} \geq p_{00}$ and $p_{01} \geq p_{00}$, then, for synergistic interactions (defined by $p_{11} \geq p'_{11}$ as given in Theorem 4), AIM has a greater difference between its interaction model and baseline model log-likelihoods than that for LR. Accordingly, since for both models the log-likelihood difference (distributed as chi-squared with the same number of degrees of freedom) is used to assess statistical significance of an interaction, the AIM model will produce strictly smaller p-values than the LR model for synergistic interactions. The proof exploits the fact that there are several ways one can determine the parameters of a logistic regression model. One way is of course to estimate the model parameters to maximize the population data log-likelihood. However, an alternative way to estimate LR parameter values is to determine them so as to be *strictly consistent with* given posterior probabilities $q_{00}$, $q_{10}$, and $q_{01}$. In particular, we note that, based on the LR model form $\log(P(C = 1|X_1, X_2)/(1 - P(C = 1|X_1, X_2))) = \beta_0 + \beta_1 X_1 + \beta_2 X_2$. Thus, for the LR model, strict consistency with $q_{00}$, $q_{10}$, and $q_{01}$ means that: $\frac{q_{00}}{1-q_{00}} = e^{\beta_0}$, $\frac{q_{10}}{1-q_{10}} = e^{\beta_0+\beta_1}$, and $\frac{q_{01}}{1-q_{01}} = e^{\beta_0+\beta_2}$. Thus, $\beta_0 = \log(\frac{q_{00}}{1-q_{00}})$, $\beta_1 = \log(\frac{q_{10}}{1-q_{10}}) - \log(\frac{q_{00}}{1-q_{00}})$, and $\beta_2 = \log(\frac{q_{01}}{1-q_{01}}) - \log(\frac{q_{00}}{1-q_{00}})$. Our proof of Theorem 4 is based on a consideration of three different LR models: i) the maximum likelihood LR model; ii) the surrogate LR model with parameters determined by the *true* posteriors, i.e. $q_{00} = p_{00}$, $q_{10} = p_{10}$, and $q_{01} = p_{01}$; iii) the surrogate LR model with parameters determined by the maximum likelihood AIM model's posteriors (denoted $p_{lm}^{(A)}$), i.e. $q_{00} = p_{00}^{(A)}$, $q_{10} = p_{10}^{(A)}$, and $q_{01} = p_{01}^{(A)}$. The proof structure is as follows. We first consider the surrogate LR model (LR') whose parameter values are determined by the (maximum likelihood) AIM model. Lemma 1 below establishes some key results concerning this surrogate LR model and the maximum likelihood AIM model. The next step is to establish a result (Lemma 2) that essentially says that a new model, formed by mixing a given model's probabilities with the true probabilities, necessarily has greater data log-likelihood than the original, given model. Finally, we exploit these Lemmas, along with the synergistic interaction assumption, to establish our (desired) detection power results. After stating Lemma 1 and Lemma 2, we proceed with the proof of Theorem 4. Proofs for the two Lemmas are given at the end of this Appendix.

*Lemma 1:* Let $p_{00}^{(A)}, p_{01}^{(A)}, p_{10}^{(A)}, p_{11}^{(A)}$ denote the posterior disease probabilities, under the four factor genotypes, for the baseline AIM model $\log(1 - P(C = 1|\underline{X})) = \beta_0' + \beta_1' X_1 + \beta_2' X_2$, where the parameter values $\beta_0', \beta_1', \beta_2'$ maximize the model's data log-likelihood on the given population. We have the following results: 1) For the LR model (denoted LR') whose parameters are determined based on $p_{00}^{(A)}, p_{01}^{(A)}, p_{10}^{(A)}$, we have $p_{00}^{(LR')} = p_{00}^{(A)}, p_{01}^{(LR')} = p_{01}^{(A)}, p_{10}^{(LR')} = p_{10}^{(A)}$; 2) *Ordering Property:* Assuming $p_{10} \geq p_{00}$, $p_{01} \geq p_{00}$, $p_{11} \geq p_{10}$, and $p_{11} \geq p_{01}$, it follows that $p_{10}^{(A)} \geq p_{00}^{(A)}$ and $p_{01}^{(A)} \geq p_{00}^{(A)}$, i.e. the MLE AIM model preserves the *ordering* of these posterior probabilities; 3) Under the same assumptions as 2), $p_{11}^{(LR')} \geq p_{11}^{(A)}$ with equality if and only if $p_{01}^{(A)} = p_{00}^{(A)}$ or $p_{10}^{(A)} = p_{00}^{(A)}$.

*Lemma 2:* Consider an $M$-category phenotype ($M \geq 2$), taking on values $\{\omega_1, \ldots, \omega_M\}$, and a population of individuals of size $N = \sum_{m=1}^{M} N_m$, $N_m$ the number of individuals possessing phenotype $\omega_m$. Let $\mathcal{Q} = \{q_m, m = 1, \ldots, M\}$ be a probability mass function model for the phenotype, and let $\mathcal{P} = \{p_m \equiv \frac{N_m}{N}, m = 1, \ldots, M\}$, i.e. it is the empirical pmf. The data log-likelihood for the population, under the model $\mathcal{Q}$, is: $L = \sum_{m=1}^{M} N_m \log(q_m) = N \sum_{m=1}^{M} p_m \log(q_m)$. Consider the new model $\mathcal{Q}' = \{\lambda p_m + (1 - \lambda) q_m, m = 1, \ldots, M\}$, where $0 \leq \lambda \leq 1$, with log-likelihood $L' = N \sum_{m=1}^{M} p_m \log(q_m')$. Then, $L' \geq L$ with equality iff $\mathcal{Q}' = \mathcal{Q}$.

We now prove Theorem 4, making use of Lemmas 1 and 2. We will only provide the proof here for the synergistic interaction case, since the proof strategy is very similar for antagonistic interactions. Let $N_{lm}, l = 0, 1\ m = 0, 1$ denote the number of subjects in the population with genotype $(X_1 = l, X_2 = m)$. Then, the log-likelihood under the baseline AIM model is: $L^{(A)} = \sum_{l=0,1} \sum_{m=0,1} N_{lm} \left( p_{lm} \log(p_{lm}^{(A)}) + (1 - p_{lm}) \log(1 - p_{lm}^{(A)}) \right)$. Likewise, we have $L^{(LR')} = \sum_{l=0,1} \sum_{m=0,1} N_{lm} \left( p_{lm} \log(p_{lm}^{(LR')}) + (1 - p_{lm}) \log(1 - p_{lm}^{(LR')}) \right)$. Now, since, from Lemma 1, $p_{00}^{(LR')} =$

$p_{00}^{(A)}, p_{01}^{(LR')} = p_{01}^{(A)}, p_{10}^{(LR')} = p_{10}^{(A)}$, we have that

$$L^{(LR')} - L^{(A)} = (N_{11}(p_{11} \log(p_{11}^{(LR')}) + (1 - p_{11}) \log(1 - p_{11}^{(LR')})) - \qquad (21)$$
$$(N_{11}(p_{11} \log(p_{11}^{(A)}) + (1 - p_{11}) \log(1 - p_{11}^{(A)}))),$$

a difference between two log-likelihoods, restricted to the subpopulation with $(X_1 = 1, X_2 = 1)$. We next consider the sign of the difference $L^{(LR')} - L^{(A)}$ under the two possible cases: $p_{11} \geq p_{11}^{(LR')}$ and $p_{11} < p_{11}^{(LR')}$.

First, suppose $p_{11} \geq p_{11}^{(LR')}$. Let $\lambda = (p_{11} - p_{11}^{(LR')})/(p_{11} - p_{11}^{(A)})$. Note that $\lambda \geq 0$ because $p_{11} \geq p_{11}^{(LR')}$ and $p_{11} \geq p_{11}^{(A)}$. Also, $\lambda < 1$ because $p_{11}^{(LR')} > p_{11}^{(A)}$ and, thus, $p_{11} - p_{11}^{(LR')} < p_{11} - p_{11}^{(A)}$. Further, one can verify that $(p_{11}^{(LR')}, 1 - p_{11}^{(LR')}) = \lambda(p_{11}^{(A)}, 1 - p_{11}^{(A)}) + (1 - \lambda)(p_{11}, 1 - p_{11})$. Thus, by Lemma 2, the log-likelihood $(N_{11}(p_{11} \log(p_{11}^{(LR')}) + (1 - p_{11}) \log(1 - p_{11}^{(LR')}))$ is greater than the log-likelihood $(N_{11}(p_{11} \log(p_{11}^{(A)}) + (1-p_{11}) \log(1-p_{11}^{(A)})))$ and, thus, $L^{(LR')} > L^{(A)}$. Finally, the maximum likelihood LR model has a log-likelihood $L^{(LR)}$ at least as large as $L^{(LR')}$, i.e. $L^{(LR)} \geq L^{(LR')} > L^{(A)}$.

Next, suppose $p_{11} < p_{11}^{(LR')}$. Let us construct the vector $\underline{p}(t) = (p_{00}(t), p_{10}(t), p_{01}(t))$, where $p_{lm}(t) = p_{lm} + t(p_{lm}^{(A)} - p_{lm})$. Note that $\underline{p}(0) = (p_{00}, p_{10}, p_{01})$ and $\underline{p}(1) = (p_{00}^{(A)}, p_{10}^{(A)}, p_{01}^{(A)})$. As shown in Lemma 1, the parameters of the LR model can be determined by these three probabilities and, thus, by the triple $\underline{p}(t)$ (for any $0 \leq t \leq 1$). Let us denote the resulting LR posterior probability, given $(X_1 = 1, X_2 = 1)$, by $p_{11}^{(LR')}(t)$, a continuous function of $t$. Now, note that $p_{11}^{(LR')}(1) = p_{11}^{(LR')} > p_{11}$. Also, $p_{11}^{(LR')}(0) = p'_{11} < p_{11}$, since this is just our definition of a synergistic interaction. We thus have that $p_{11}^{(LR')}(0) < p_{11} < p_{11}^{(LR')}(1)$. Since $p_{11}^{(LR')}(t)$ is a continuous function, by the intermediate value theorem, there must be some value $t_c$, $0 < t_c < 1$, such that $p_{11}^{(LR')}(t_c) = p_{11}$. Let us consider the log-likelihood for this model, $L^{(LR')}(t_c)$, which can be written as $L^{(LR')}(t_c) = \sum_{l=0,1} \sum_{m=0,1} L_{lm}^{(LR')}(t_c)$, where $L_{lm}^{(LR')}(t_c) = N_{lm}(p_{lm} \log(p_{lm}^{(LR')}(t_c)) + (1-p_{lm}) \log(1-p_{lm}^{(LR')}(t_c)))$. Now, for $(l, m) = (0, 0), (0, 1),$ and $(1, 0)$, we have that $(p_{lm}^{(LR')}(t_c), 1 - p_{lm}^{(LR')}(t_c)) = t_c(p_{lm}^{(A)}, 1 - p_{lm}^{(A)}) + (1 - t_c)(p_{lm}, 1 - p_{lm})$. Recalling that $0 < t_c < 1$ and applying Lemma 2, we have that $L_{lm}^{(LR')}(t_c) > L_{lm}^{(A)}$, $(l, m) = (0, 0), (1, 0), (0, 1)$, where $L_{lm}^{(A)}$ is the log-likelihood for the $(X_1 = l, X_2 = m)$ subpopulation under the AIM posterior model. Finally, since $p_{11}^{(LR')}(t_c) = p_{11}$, we have $(p_{11}^{(LR')}(t_c), 1-p_{11}^{(LR')}(t_c)) = \lambda(p_{11}, 1 - p_{11}) + (1 - \lambda)(p_{11}^{(A)}, 1 - p_{11}^{(A)})$, where $\lambda = 1$. Thus, again applying Lemma 2, we have: $L_{11}^{(LR')}(t_c) > L_{11}^{(A)}$.

Since all four genotype-conditioned log-likelihoods for the LR model determined based on $\underline{p}(t_c)$ are greater than their counterpart log-likelihoods for the maximum likelihood AIM model, we have, for the composite log-likelihoods, that $L^{(LR')}(t_c) > L^{(A)}$. Moreover, the maximum likelihood LR model has log-likelihood $L^{(LR)} \geq L^{(LR')}(t_c)$. Thus, we again obtain $L^{(LR)} > L^{(A)}$.

In summary, under both possible scenarios ($p_{11} \geq p_{11}^{(LR')}$ and $p_{11} < p_{11}^{(LR')}$), $L^{(LR)} > L^{(A)}$. Finally, for hypothesis testing, we look at the difference between the log-likelihood under the alternative hypothesis (where an interaction term $\beta_3 X_1 X_2$ is included in the model), $L_{Alt}$, and the baseline model log-likelihood. Now, under the alternative hypothesis, for *both* AIM and LR, the models are *saturated*, i.e. there are four genotype values and four free parameters. Thus, the AIM and LR alternative hypothesis maximum likelihood models have the same log-likelihood value $L_{alt}$ (Hosmer(2013)), which must be at least as large as the baseline model log-likelihoods. Thus, $L_{alt} \geq L^{(LR)}$, and since $L^{(LR)} > L^{(A)}$, we have $L_{alt} - L^{(LR)} < L_{alt} - L^{(A)}$, i.e. AIM gives a greater log-likelihood difference.

Q.E.D.

*Proof of Lemma 1:*

**First Result:** Under the AIM model, $p_{00}^{(A)} = 1 - e^{\beta_0'}$, $p_{10}^{(A)} = 1 - e^{\beta_0' + \beta_1'}$, and $p_{01}^{(A)} = 1 - e^{\beta_0' + \beta_2'}$. Now, consider the LR model whose parameters are determined based on the AIM model's posterior probabilities $p_{00}^{(A)}, p_{10}^{(A)}, p_{01}^{(A)}$, rather than based on the true disease posteriors $p_{00}, p_{10}, p_{01}$. Accordingly, we let $\beta_0 = \log(\frac{p_{00}^{(A)}}{1-p_{00}^{(A)}})$, $\beta_1 = \log(\frac{p_{10}^{(A)}}{1-p_{10}^{(A)}}) - \log(\frac{p_{00}^{(A)}}{1-p_{00}^{(A)}})$, and $\beta_2 = \log(\frac{p_{01}^{(A)}}{1-p_{01}^{(A)}}) - \log(\frac{p_{00}^{(A)}}{1-p_{00}^{(A)}})$. Based on these parameter value assignments, the LR posterior probabilities are:

$$p_{00}^{(LR')} = e^{\beta_0}/(1+e^{\beta_0})\Big|_{\beta_0 = \log(\frac{p_{00}^{(A)}}{1-p_{00}^{(A)}})} = p_{00}^{(A)} \quad (22)$$

$$p_{10}^{(LR')} = e^{\beta_0+\beta_1}/(1+e^{\beta_0+\beta_1})\Big|_{\beta_0=\log(\frac{p_{00}^{(A)}}{1-p_{00}^{(A)}}), \beta_1=\log(\frac{p_{10}^{(A)}}{1-p_{10}^{(A)}})-\log(\frac{p_{00}^{(A)}}{1-p_{00}^{(A)}})} = p_{10}^{(A)}.$$

Likewise, it is also found that $p_{01}^{(LR')} = p_{01}^{(A)}$.

**Second Result:**

The log-likelihood under an AIM model is:

$$L = \sum_{l=0,1} \sum_{m=0,1} p_{lm} \log(p_{lm}^{(A)}), \quad (23)$$

where $p_{00}^{(A)} = 1-e^{\beta_0}$, $p_{10}^{(A)} = 1-e^{\beta_0+\beta_1}$, $p_{01}^{(A)} = 1-e^{\beta_0+\beta_2}$, and $p_{11}^{(A)} = 1-e^{\beta_0+\beta_1+\beta_2}$. The maximum likelihood AIM model satisfies the necessary optimality conditions: $\partial L/\partial \beta_i = 0, i = 0, 1, 2$. Taking derivatives, we find that these conditions are:

$$\frac{\partial L}{\partial \beta_0} = \sum_{l=0,1} \sum_{m=0,1} N_{lm} \left( \frac{p_{lm}^{(A)} - p_{lm}}{p_{lm}^{(A)}} \right) = 0 \quad (24)$$

$$\frac{\partial L}{\partial \beta_1} = N_{10}\left(\frac{p_{10}^{(A)} - p_{10}}{p_{10}^{(A)}}\right) + N_{11}\left(\frac{p_{11}^{(A)} - p_{11}}{p_{11}^{(A)}}\right) = 0$$

$$\frac{\partial L}{\partial \beta_2} = N_{01}\left(\frac{p_{01}^{(A)} - p_{01}}{p_{01}^{(A)}}\right) + N_{11}\left(\frac{p_{11}^{(A)} - p_{11}}{p_{11}^{(A)}}\right) = 0.$$

Now, note that the term $N_{11}\left(\frac{p_{11}^{(A)} - p_{11}}{p_{11}^{(A)}}\right)$ is common to the $\beta_1$ and $\beta_2$ derivative conditions. Thus, we have that $N_{10}\left(\frac{p_{10}^{(A)} - p_{10}}{p_{10}^{(A)}}\right) = N_{01}\left(\frac{p_{01}^{(A)} - p_{01}}{p_{01}^{(A)}}\right) = -N_{11}\left(\frac{p_{11}^{(A)} - p_{11}}{p_{11}^{(A)}}\right)$. Further, since $N_{lm} > 0$ and $p_{lm}^{(A)} > 0 \forall l, m$, this equality implies two possible cases: 1) $p_{10}^{(A)} \geq p_{10}$, $p_{01}^{(A)} \geq p_{01}$, and $p_{11}^{(A)} \leq p_{11}$; 2) $p_{10}^{(A)} \leq p_{10}$, $p_{01}^{(A)} \leq p_{01}$, and $p_{11}^{(A)} \geq p_{11}$.

Assuming the first case, we then have $B \equiv N_{10}\left(\frac{p_{10}^{(A)} - p_{10}}{p_{10}^{(A)}}\right) = N_{01}\left(\frac{p_{01}^{(A)} - p_{01}}{p_{01}^{(A)}}\right) = -N_{11}\left(\frac{p_{11}^{(A)} - p_{11}}{p_{11}^{(A)}}\right) \geq 0$. Thus, for this case, the derivative condition for $\beta_0$ can be re-expressed as: $\frac{\partial L}{\partial \beta_0} = \left(\frac{p_{00}^{(A)} - p_{00}}{p_{00}^{(A)}}\right) + B + B - B = 0$, $B \geq 0$. Equality can only be satisfied if the first term is non-positive, which requires $p_{00}^{(A)} \leq p_{00}$. However, since $p_{10} \geq p_{00}$ and $p_{01} \geq p_{00}$, this implies $p_{10}^{(A)} \geq p_{00}^{(A)}$ and $p_{01}^{(A)} \geq p_{00}^{(A)}$.

Next, consider the second case. Suppose that $p_{10}^{(A)} < p_{00}^{(A)}$. This implies that $\beta_1 > 0$, which implies that $p_{11}^{(A)} < p_{10}^{(A)}$. Moreover, $p_{10}^{(A)} \leq p_{10}$. Thus, $p_{11}^{(A)} < p_{10}^{(A)} \leq p_{10} \leq p_{11}$. However, under this case we also have that $p_{11}^{(A)} \geq p_{11}$. Thus, the assumption that $p_{10}^{(A)} < p_{00}^{(A)}$ leads to a contradiction. Applying the same logic, one can show, for this case, that $p_{01}^{(A)} < p_{00}^{(A)}$ also leads to contradiction. Thus, for this case, we must have $p_{10}^{(A)} \geq p_{00}^{(A)}$ and $p_{01}^{(A)} \geq p_{00}^{(A)}$.

Under both possible cases (and thus, in general), we have: $p_{10}^{(A)} \geq p_{00}^{(A)}$ and $p_{01}^{(A)} \geq p_{00}^{(A)}$. Q.E.D.

**Third Result:**

$$\frac{p_{11}^{(LR')}}{1 - p_{11}^{(LR')}} = e^{\beta_0 + \beta_1 + \beta_2} = \frac{e^{\beta_0 + \beta_1} e^{\beta_0 + \beta_2}}{e^{\beta_0}} \tag{25}$$

$$= \frac{(p_{10}^{(LR')}/(1 - p_{10}^{(LR')}))(p_{01}^{(LR')}/(1 - p_{01}^{(LR')}))}{(p_{00}^{(LR')}/(1 - p_{00}^{(LR')}))}.$$

With simple algebra, we obtain:

$$1 - p_{11}^{(LR')} = \frac{(1 - p_{01}^{(LR')})(1 - p_{10}^{(LR')})p_{00}^{(LR')}}{p_{01}^{(LR')} p_{10}^{(LR')}(1 - p_{00}^{(LR')})} p_{11}^{(LR')} \tag{26}$$

$$= \frac{(1 - p_{01}^{(A)})(1 - p_{10}^{(A)})p_{00}^{(A)}}{p_{01}^{(A)} p_{10}^{(A)}(1 - p_{00}^{(A)})} p_{11}^{(LR')}.$$

Solving for $p_{11}^{(LR')}$, we then obtain:

$$p_{11}^{(LR')} = \frac{p_{01}^{(A)} p_{10}^{(A)}(1 - p_{00}^{(A)})}{p_{00}^{(A)} - p_{01}^{(A)} p_{00}^{(A)} - p_{10}^{(A)} p_{00}^{(A)} + p_{01}^{(A)} p_{10}^{(A)}}. \tag{27}$$

Correspondingly, for the maximum likelihood model of AIM form $\log(1 - P(C = 1|X_1, X_2))) = \beta'_0 + \beta'_1 X_1 + \beta'_2 X_2$, we have $\beta'_0 = \log(1 - p^{(A)}_{00})$, $\beta'_1 = \log(\frac{1-p^{(A)}_{10}}{1-p^{(A)}_{00}})$, and $\beta'_2 = \log(\frac{1-p^{(A)}_{01}}{1-p^{(A)}_{00}})$. Thus,

$$1 - p^{(A)}_{11} = e^{\beta'_0 + \beta'_1 + \beta'_2} = \frac{(1 - p^{(A)}_{10})(1 - p^{(A)}_{01})}{(1 - p^{(A)}_{00})} \text{ and} \tag{28}$$

$$p^{(A)}_{11} = 1 - \frac{(1 - p^{(A)}_{10})(1 - p^{(A)}_{01})}{(1 - p^{(A)}_{00})}.$$

Now, let us check the sign of $p^{(LR')}_{11} - p^{(A)}_{11}$. We can write:

$$\begin{aligned} p^{(LR')}_{11} - p^{(A)}_{11} &= (1 - p^{(LR')}_{11})\left(\frac{(1 - p^{(A)}_{11})}{(1 - p^{(LR')}_{11})} - 1\right) \\ &= (1 - p^{(LR')}_{11})\left(\frac{p^{(A)}_{01} p^{(A)}_{10}}{p^{(A)}_{00} p^{(LR')}_{11}} - 1\right), \end{aligned} \tag{29}$$

where the latter expression is obtained using (26) and (28).

We now compare $\frac{p^{(A)}_{01} p^{(A)}_{10}}{p^{(A)}_{00} p^{(LR')}_{11}}$ to 1. First, using (27), we have:

$$\begin{aligned} \frac{p^{(A)}_{01} p^{(A)}_{10}}{p^{(A)}_{00} p^{(LR')}_{11}} &= \frac{p^{(A)}_{00} - p^{(A)}_{00} p^{(A)}_{01} - p^{(A)}_{00} p^{(A)}_{10} + p^{(A)}_{01} p^{(A)}_{10}}{p^{(A)}_{00}(1 - p^{(A)}_{00})} \\ &= \frac{p^{(A)}_{01}(p^{(A)}_{10} - p^{(A)}_{00}) + p^{(A)}_{00} - p^{(A)}_{00} p^{(A)}_{10}}{p^{(A)}_{00}(1 - p^{(A)}_{00})} \\ &= \frac{(p^{(A)}_{01} - p^{(A)}_{00})(p^{(A)}_{10} - p^{(A)}_{00}) + p^{(A)}_{00}(p^{(A)}_{10} - p^{(A)}_{00}) + p^{(A)}_{00} - p^{(A)}_{00} p^{(A)}_{10}}{p^{(A)}_{00}(1 - p^{(A)}_{00})} \\ &= \frac{(p^{(A)}_{01} - p^{(A)}_{00})(p^{(A)}_{10} - p^{(A)}_{00}) + p^{(A)}_{00}(1 - p^{(A)}_{00})}{p^{(A)}_{00}(1 - p^{(A)}_{00})}. \end{aligned} \tag{30}$$

Now, since $(p^{(A)}_{01} - p^{(A)}_{00}) \geq 0$ and $(p^{(A)}_{10} - p^{(A)}_{00}) \geq 0$, the final expression must also be greater than or equal to 1. Thus, we have $\frac{p^{(A)}_{01} p^{(A)}_{10}}{p^{(A)}_{00} p^{(LR')}_{11}} \geq 1$. Now, examining (29) and noting that $1 - p^{(LR')}_{11} > 0$, we have finally proved that $p^{(LR')}_{11} \geq p^{(A)}_{11}$. Furthermore, examining $(p^{(A)}_{01} - p^{(A)}_{00})(p^{(A)}_{10} - p^{(A)}_{00})$ in the final expression in (30), it is seen that this expression equals 1, and, thus, $p^{(LR')}_{11} = p^{(A)}_{11}$, if and only if $p^{(A)}_{01} = p^{(A)}_{00}$ or $p^{(A)}_{10} = p^{(A)}_{00}$.

Q.E.D.

*Proof of Lemma 2:*

$$L' - L = N \sum_{m=1}^{M} p_m \log(q'_m) - N \sum_{m=1}^{M} p_m \log(q_m) \tag{31}$$

$$\begin{aligned}
&= N \sum_{m=1}^{M} p_m \log(\lambda p_m + (1-\lambda) q_m) - N \sum_{m=1}^{M} p_m \log(q_m) \\
&\geq \sum_{m=1}^{M} (\lambda p_m \log(p_m) + (1-\lambda) p_m \log(q_m)) - \sum_{m=1}^{M} p_m \log(q_m) \\
&= \lambda \sum_{m=1}^{M} p_m \log(\frac{p_m}{q_m}) = \lambda D_{\text{KL}}(\mathcal{P}||\mathcal{Q}) \geq 0.
\end{aligned}$$

Here, the first inequality is obtained from Jensen's inequality applied to the logarithm function, and $D_{\text{KL}}(\mathcal{P}||\mathcal{Q})$ is the Kullback-Leibler distance between pmfs (which is non-negative). Thus, $L' \geq L$. Note that equality is achieved if $\lambda = 0$, in which case $q'_m = q_m \forall m$. Moreover, since the $\log()$ is strictly concave, again by Jensen's inequality, equality is *only* possible for $\lambda = 0$ or $1$; in this case, only if $q'_m = q_m \forall m$.

Q.E.D.